\documentstyle[graphics]{cupconf}

\ifoldfss
\else
  \ifnfssone
    \newmathalphabet{\mathit}
      \addtoversion{normal}{\mathit}{cmr}{m}{it}
      \addtoversion{bold}{\mathit}{cmr}{bx}{it}
    \newmathalphabet{\mathcal}
      \addtoversion{normal}{\mathcal}{cmsy}{m}{n}
    \else
    \ifnfsstwo
    \fi
  \fi
\fi

\def\gtrsim{\mathrel{\hbox{\rlap{\hbox{\lower4pt\hbox{$\sim$}}}\hbox{$>$}}}}
\def\lesssim{\mathrel{\hbox{\rlap{\hbox{\lower4pt\hbox{$\sim$}}}\hbox{$<$}}}}

\def\hexnumber#1{\ifcase#1 0\or1\or2\or3\or4\or5\or6\or7\or8\or9\or
 A\or B\or C\or D\or E\or F\fi }

\makeatletter
\ifx\CUP@mtlplain@loaded\undefined
\else
\fi
\makeatother
 \makeatletter
 \ifx\CUP@mtlplain@loaded\undefined
   \font\tenbmi=cmmib10 at 10pt
   \font\sevenbmi=cmmib10 at 7pt
   \font\fivebmi=cmmib10 at 5pt

   \newfam\bmifam
   \textfont\bmifam=\tenbmi
   \scriptfont\bmifam=\sevenbmi
   \scriptscriptfont\bmifam=\fivebmi
   
 \fi
 \makeatother
\ifnfsstwo

\fi
\ifnfssone

\fi
\ifoldfss

\fi

\mathchardef\varLambda="0103

\makeatletter
\ifx\CUP@mtlplain@loaded\undefined
\else
\fi
\makeatother
\makeatletter
\ifx\CUP@mtlplain@loaded\undefined
  \font\tenbms=cmbsy10
  \font\sevenbms=cmbsy10 at 7pt
  \font\fivebms=cmbsy10 at 5pt
  \newfam\bmsfam
  \textfont\bmsfam=\tenbms
  \scriptfont\bmsfam=\sevenbms
  \scriptscriptfont\bmsfam=\fivebms

  \edef\bsy@{\hexnumber\bmsfam}
  \mathchardef\bnabla="0\bsy@72
\fi
\makeatother
\begin{document}
\title[Chemical Evolution of Galaxies and Intracluster Medium]{Chemical Evolution of Galaxies and Intracluster Medium}

\author[Francesca Matteucci]%
{F\ls R\ls A\ls N\ls C\ls E\ls S\ls C\ls A\ns M\ls A\ls T\ls T\ls
 E\ls U\ls C\ls C\ls I$^1$}

\affiliation{$^1$
Department of Astronomy, University of Trieste\\
Via G.B. Tiepolo 11\\
34100 Trieste, Italy}

\setcounter{page}{1}


\ifnfssone
\else
  \ifnfsstwo
  \else
    \ifoldfss
      \let\mathcal\cal
      \let\mathrm\rm
      \let\mathsf\sf
    \fi
  \fi
\fi

\maketitle

\begin{abstract}
In this series of lectures I discuss the basic principles 
and the modelling of the chemical evolution of galaxies. In particular, I
present models for the chemical evolution of the Milky Way galaxy 
and compare them with the available observational data. 
From this comparison one can infer important constraints 
on the mechanism of formation of the Milky Way as well as
on stellar nucleosynthesis and supernova progenitors.
Models for the chemical evolution of elliptical galaxies are also 
shown in the framework of the two competing scenarios for galaxy formation: 
monolithic and hierachical.
The evolution of dwarf starbursting galaxies is also presented and 
the connection of these objects with Damped Lyman- $\alpha$ systems is 
briefly discussed.
The roles of supernovae of different type (I, II) is discussed in 
the general framework of galactic evolution and in connection with 
the interpretation of high redshift objects.
Finally, the chemical enrichment of the intracluster medium as due 
mainly to ellipticals and S0 galaxies is discussed. 
\end{abstract}

\section{Basic parameters of chemical evolution}
Galactic chemical evolution is the study of the evolution in time and space 
of the abundances of the chemical elements in the interstellar gas in 
galaxies. This process is influenced by many parameters such as the
initial conditions, the 
star formation and evolution, the nucleosynthesis and possible gas flows.

Here I describe each one separately:
\begin{itemize}
\item Initial conditions- 
One can assume that all the initial gas out of which the galaxy will form is
already present when the star formation process starts or that the gas
is slowly accumulated in time. Then one can assume that the initial 
chemical composition of this gas is primordial (no metals) or that  some pre-
enrichment has already taken place (e.g. Population III stars). 
As we will see in the following, different assumptions are required 
for different galaxies.

\item The birthrate function- Stars form and die continuously in galaxies, 
therefore a recipe for star formation is necessary.
We define the stellar birthrate function as the number of stars formed 
in the time interval $dt$ and in the mass range $dm$ as:
\begin{equation}
B(m,t)=\psi(t) \varphi(m)dtdm
\end{equation}
where:
\begin{equation}
\psi(t)=SFR
\end{equation}
is the star formation rate (SFR), and:
\begin{equation}
\varphi(m)=IMF
\end{equation}
is the initial mass function (IMF). The SFR is assumed to be only a function 
of time and the IMF only a function of mass. This is clearly an 
oversemplification but is necessary in absence of a clear knowledge 
of the star formation process.

\item Stellar evolution and nucleosynthesis- Nuclear burnings take place 
in the stellar interiors during the star lifetime and produce new chemical 
elements, in particular metals. 
These metals, together with the pristine stellar material is restored 
into the interstellar medium  (ISM) at the star death.
This process 
clearly affects crucially the chemical evolution of the ISM.
In order to take into account the elemental production by stars we define
the ``yields'', in particular the {\it stellar yields} (the amount of 
elements produced by a single star) and the {\it yields per stellar 
generation} (the amount
of elements produced by an entire stellar generation).

\item Supplementary Parameters-
Infall of extragalactic gas, radial flows and galactic winds are
important ingredients in building galactic 
chemical evolution models.
\end{itemize}

\section{The stellar birthrate}
\subsection{Theoretical recipes for the SFR}
Several parametrizations, besides the simple one of a constant $\psi(t)$, 
are used in the literature for the SFR:
\begin{itemize}
\item Exponentially decreasing:
\begin{equation}
SFR= \nu e^{-t/ \tau_*}
\end{equation}
with $\tau_* = 5-15$ Gyr in order to produce realistic values
which can be compared with the present time SFR in the Milky Way (Tosi, 1988).

\item The Schmidt (1965) law:
\begin{equation}
SFR = \nu \sigma_{gas}^{k}
\end{equation}
is the most widely adopted formulation for the SFR. It was originally 
formulated by Schmidt (1959;1963) as a function of the volume gas density 
with $k=2.0$. He measured the space density of stars in different regions 
of the Galaxy in relation to the number density of neutral hydrogen, 
measured by 
means of the 21 cm emission
The formulation as a function of the surface gas density $\sigma_{gas}$
is normally preferred for studying the Milky Way disk and galactic disks 
in general. In principle, Schmidt's formulation
as a function of the volume gas density and that with $\sigma_{gas}$ 
are equivalent when $k=1$.
Kennicutt, in a series of papers (1983;1989;1998a,b) tried to assess the 
dependence of massive star formation on the surface gas density in disk galaxies, 
by comparing $H_{\alpha}$ emission with the data on the distribution of 
HI and CO and found that the SFR can be well represented by 
a Schmidt law with 
$k=1.4 \pm0.15$. The quantity $\nu$ is the efficiency of star formation and is expressed in units of $time^{-1}$.
\item A more complex formulation, which depends also upon the total 
surface mass density, was suggested by Dopita and Ryder (1994) 
and can be written as:
\begin{equation}
SFR= \nu \sigma_{tot}^{k_1} \sigma_{gas}^{k_2}
\end{equation}
with $\sigma_{tot}$ being the total surface mass density. This kind of 
SFR is related to the feedback mechanism between the energy injected 
into the ISM by supernovae (SNe) and stellar 
winds and the local potential well.

\item In alternative to the Schmidt law, Kennicutt proposed also the 
following star formation law, which fits equally well the observational data:
\begin{equation}
SFR= 0.017 \Omega_{gas} \sigma_{gas}\propto R^{-1} \sigma_{gas}
\end{equation}
with $\Omega_{gas}$ being the angular rotation speed of 
the gas.
\end{itemize}

\subsection {The tracers of star formation}
The main tracers of star formation in galaxies are:
\begin{itemize} 
\item Counts of
luminous supergiants 
in nearby galaxies under the assumption that 
their number is proportional to the SFR. 

\item 
The $H_{\alpha}$ and $H_{\beta}$ flux 
from HII regions, which are ionized by young 
and hot stars, under the assumption that such flux is proportional to the SFR
(Kennicutt 1998): 
\begin{equation}
SFR(M_{\odot}yr^{-1})={7.9 \cdot 10^{-42} L_{H_{\alpha}}(erg s^{-1})} 
\end{equation}

\item From the integrated UBV colors and spectra of galaxies one 
can estimate the relative proportions of 
young and old stars and derive the ratio
between the present time SFR and the average SFR in the past.

\item 
The frequency of type II SNe as well as the distribution of SN remnants 
and pulsars can be used as tracers 
of the SFR. These tracers have been used for deriving the SFR in the 
Galactic disk.

\item
The radio emission
from HII regions can also be a tracer of the SFR.

\item
The ultraviolet continuum and the infrared continuum 
(star forming regions are surrounded by dust) are also connected to the SFR
as in the following expression:

\begin{equation}
SFR(M_{\odot} yr^{-1})= 0.9 \cdot 10^{-6} {L(UV) \over L_{bol_{\odot}}}
\end{equation}
derived by means of a two-slope IMF by Donas et al. (1987).

\item 
Finally, the SFR can be derived from the distribution of molecular clouds
(Rana 1991).
\end{itemize}

All of these formulations for the SFR need the assumption of an IMF and 
viceversa, the derivation of the IMF needs the assumption of a star 
formation history, as described in the next section.
The local SFR, derived under the assumption of a particular IMF,
suitable for the solar neighbourhood, gives the following range of values:
\begin{equation}
SFR=2-10 \rm M_{\odot}pc^{-2} Gyr^{-1} 
\end{equation}
(see Timmes et al. 1995)

\subsection{The IMF: Various Parametrizations}

The IMF is a probability distribution function 
and is normally approximated by a power law, namely:

\begin{equation}
\varphi(m)=am^{-(1+x)}
\end{equation}
which is the number of stars with masses in the interval m, m+dm. 
The IMF can be one-slope
(Salpeter (1955) x=1.35) or multi-slope
(Scalo 1986,1998; Kroupa et al. 1993).
The IMF is usually normalized as:
\begin{equation}
\int^{\infty}_{0}m{\varphi(m)dm}=1
\end{equation}
The IMF is derived locally 
from the present day mass function (PDMF)
which in turn is obtained by counting the Main Sequence stars per 
interval of magnitude. Then the star counts are transformed into number 
of stars per $pc^{2}$  and then a mass-luminosity relation is adopted 
to pass from the luminosity to the mass.

\subsection{Derivation of the IMF}
As already mentioned, the IMF is derived by the observed PDMF, which is
the current mass distribution of Main-Sequence (MS) stars  
per unit area,
$n(m)$.

For stars ($0.1 < m/m_{\odot} \le 1$)
with lifetimes $\tau_m \ge t_G$ (with $t_G$ being the Galactic age), 
$n(m)$ can be written as:

\begin{equation}
n(m)=\int^{t_G}_{0}{\varphi(m) \psi(t) dt}
\end{equation}
where $t_G$ is the Galactic lifetime. These stars
are all still on the MS.
If $\varphi(m)$ is constant in time, as it is usually assumed, then: 
\begin{equation}
n(m) =\varphi(m) <\psi> t_G  
\end{equation}
where
$<\psi>$ is the average SFR in the past.

For stars with 
$\tau_m << t_G$ ($m \ge 2 M_{\odot}$), we see on the MS only those born after
the time ($t=t_{G}-\tau_m$). 
The PDMF is therefore: 
\begin{equation}
n(m)= \int^{t_G}_{t_G-\tau_m}{\varphi(m) \psi(t)dt}
\end{equation}

Again,
if $\varphi(m)$ is constant in time:
\begin{equation}
n(m)=\varphi(m) \psi(t_G) \tau_{m}
\end{equation}
under the assumption that $\psi(t_{G})=\psi(t_G- \tau_{m})$, where 
$\psi(t_G)$ is the  SFR at the present time, $t_G$, and $\tau_m$ is the
lifetime of a star of mass $m$.

The IMF, $\varphi(m)$, in the mass interval $1-2 ~M_{\odot}$ depends on
the ratio:
\begin{equation}
b(t_G)={\psi(t_G) \over <\psi>}
\end{equation}
It has been shown by Scalo (1986) that a good fit between the 
two portions of the $\varphi(m)$,
namely below $1M_{\odot}$ and above $2M_{\odot}$, requires:
\begin{equation}
0.5 \le b(t_G) \le 1.5
\end{equation}
which means that the SFR in the local disk should have varied in time 
less than a factor of 2 during the whole disk lifetime.

\subsection{The Infall Rate: Various Parametrizations}
The presence of infall of extragalactic gas in the chemical evolution of 
galaxies is demanded by the G-dwarf metallicity distribution in the solar 
vicinity
and by the existence of high velocity clouds infalling towards 
the galactic disk. The origin of this infalling gas on the Galaxy is not 
yet entirely clear and measurements of the metallicity of such gas are 
necessary to decide whether it originates in the galactic disk (galactic 
fountain) or if it has an extragalactic origin.

Several parametrizations for the infall rate have been used so far:
\begin{itemize}
\item Constant in space and time

\item Variable in space and time such as:
\begin{equation}
IR= A(R) e^{-t/ \tau(R)}
\end{equation}
with $\tau(R)$ is constant or varying along the disk and
$A(R)$ is derived by fitting the present time distribution of
$\sigma_{tot}(R,t_{G})$. 

\end{itemize}

\section {Nucleosynthesis}
During the Big Bang the light elements 
(D, $^{3}He$, $^{4}He$ and $^{7}Li$) were produced. 
On the other hand, all the elements heavier than $^{7}Li$,
with the exception of Be and B, are produced inside stars.
The light elements $^{6}Li$, Be and B are instead manufactured
by spallation processes 
in the ISM due to the interaction between cosmic rays 
and interstellar atoms. 
\subsection{Nucleosynthesis in the Big Bang}
I give here a brief summary of the main steps in the Big Bang nucleosynthesis.

When the temperature in the universe was $T=10^{12}$K, 
only weak interactions causing conversions between 
protons and neutrons occurred, namely:
\begin{equation}
p + e \leftrightarrow n + \nu,
\end{equation}

\begin{equation}
p + \tilde \nu \leftrightarrow  n + e^{+}.
\end{equation}

The nucleosynthesis started when $T=10^{9}$K and 
lasted until $T=10^{8}$K. The first element to be formed was  
D (a nucleus composed by a neutron plus a proton)
and subsequently $^{3}He$ from the reaction:

\begin{equation}
D + p \rightarrow ^{3}He + \gamma
\end{equation}
followed by:

\begin{equation}
^{3}He + n \rightarrow ^{4}He + \gamma
\end{equation}
Then also very small fractions of $^{7}Li$ ($10^{-9}$ by mass)
and $^{7}Be$ ($10^{-11}$ by mass) were produced.

One of the major achievements in cosmology is that it can account
simultaneously for the primordial abundances of H, D, $^{3}He$, $^{4}He$ 
and $^{7}Li$ but only for a low density universe. The comparison between 
the observed primordial abundances and the Big Bang nucleosynthesis 
calculations
can allow to impose constraints upon the baryon to photon ratio ($\eta$)
in the universe.
In particular, for a baryon to photon ratio 
$\eta \sim 3 \cdot 10^{-10}$
the baryonic density parameter of the universe is (Peacock, 1999):
\begin{equation}
0.010 \le \Omega_b h^{2} \le 0.015
\end{equation}

\subsection{Stellar Nucleosynthesis}
Before discussing stellar nucleosynthesis we need to define the 
crucial stellar mass ranges.

\begin{itemize}
\item Brown Dwarfs ($M < M_L$,  $M_{L}=0.08-0.09M_{\odot}$).
They never ignite H and their lifetimes are larger than the age of the 
universe.

\item Low mass stars ($0.5 \le M/M_{\odot} \le M_{HeF}$) 
($M_{HeF}$=1.85-2.2$M_{\odot}$ depending on stellar models)
ignite He explosively and become C-O  white dwarfs (WD).
If $M < 0.5 M_{\odot}$
they become He WD. The lifetimes range from several $10^{9}$ years 
up to several Hubble times.

\item Intermediate mass stars 
($M_{HeF} \le M/M_{\odot} \le M_{up}$) ignite He quiescently. 
$M_{up}$ is the limiting mass for the formation of a C-O degenerate core
and is in the range 5-9$M_{\odot}$, depending on stellar evolution models.
Their lifetimes range from several $10^{7}$ to  $10^{9}$ years.
These stars die as C-O WDs if they are not in binary systems.
If in binary systems, stars in this mass range can give rise to Type Ia SNe
(see later).

\item Massive stars ($M >M_{up}$)\par

Stars in the mass range $M_{up} \le M/M_{\odot} \le 10-12$ 
become e-capture SNe 
(Type II SNe) and leave
neutron stars as remnants.
Stars in the range 
$10-12 \le M/M_{\odot} \le M_{WR}$ ($M_{WR} \sim 20-40 M_{\odot}$)
end their lives as core-collapse SNe  (Type II) and leave
a neutron star or a black hole
as a remnant.
Stars in the range $M_{WR} \le M/M_{\odot} \le 100$ probably
become Type Ib SNe.
The lifetimes of these stars are in the range from several
$10^{7}$ to $\sim 10^{6}$ years.

\item  Very Massive Stars ($M > 100 M_{\odot}$), if they exist, 
explode by means of ``pair creation'' and are called
pair-creation SNe. In fact, at 
$T \sim 2 \cdot 10^{9}$ K a large portion of the gravitational energy 
goes into creation of pairs $(e^{+}, e^{-})$, thus the star 
becomes unstable and explodes.
These SNe leave no remnant and have 
lifetimes $< 10^{6}$ years.

\item Supermassive objects 
($400 \le M/M_{\odot} \le 7.5 \cdot 10^{5}$), 
if they exist, either explode due to explosive H-burning or collapse 
directly to black
holes. The only available nucleosynthesis calculations 
for these stars are from Woosley et al. (1984).

\end{itemize}

All the elements with mass number $A$ from 12 to 60 have 
been formed in stars during
the quiescent burnings occurring during their lifetime.
The main nuclear burnings are H, He, C, Ne, O and Si.
Stars transform H into He and then He into heaviers until the 
Fe-peak elements, where the binding energy per nucleon reaches a maximum.
At this point nuclear fusion reactions cannot occur anymore and the Fe nucleus
starts contracting. When the central density reaches the atomic nuclear 
density, matter becomes uncompressible and a core-bounce occurs with the 
consequent formation of a shock wave and ejection of the star mantle.
However, a problem exists for the explosion of stars with large Fe cores since
most of the collapse gravitational energy is used to photodisintegrate Fe,
thus weakening the shock wave and preventing the mantle ejection. 
To overcome this problem it has been suggested the existence of some 
mechanisms 
able to rejuvenate the shock wave, such as
neutrino-heating from the collapsing neutron star, rotation and 
magnetic fields.
 
Here is a summary of the main nucleosynthesis stages:
the first element to be burned is H, which
is transformed into He through the proton-proton 
chain or the 
CNO-cycle, then $^{4}He$ is transformed into $^{12}C$ through the 
triple $\alpha$
reaction.
Elements heavier than $^{12}C$ are then produced by synthesis 
of $\alpha$-particles  thus producing the
so-called $\alpha$-elements (O, Ne, Mg, Si, S, Ca and Ti).
The last main burning in stars is the $^{28}Si$ -burning which produces
$^{56}Ni$ which then $\beta$-decays into $^{56}Co$ and $^{56}Fe$.
Si-burning can be quiescent or explosive (depending on the temperature)
but it always produces Fe.
Explosive nucleosynthesis  occurs in the inverse order 
(Si, O, Ne, C, He, H) relative
to quiescent nucleosynthesis, depending on the fact that it starts 
in the center and propagates outwards following the passage of the shock wave.
The main products of explosive nucleosynthesis are the Fe-peak elements.
It is worth noting that low and intermediate mass stars never ignite 
C and thus end their lives as C-O white dwarfs.
Stars with masses below 10-- 12 $M_{\odot}$ 
(depending on stellar models) explode during O-burning and end up as 
e-capture SNe. Only massive stars can ignite all six nuclear fuels 
until they 
form an Fe-core.
S- and r-process elements (elements with A$> 60$ up to Th and U)
are formed by means of slow or rapid (relative to the $\beta$- decay)
neutron capture by Fe seed nuclei.
In particular, s-processing occurs during quiescent He-burning both 
in massive and
low and intermediate mass stars,
whereas r-processing
occurs during SN explosions. 

\subsection{Supernova Progenitors}

Supernovae, planetary nebulae (PNe) and, to a minor extent, stellar winds 
are the means to 
restore
the nuclearly enriched material into the ISM, thus giving rise to the process 
of chemical evolution.
There are two main Types of SNe (II, I) then divided in subclasses:
SNe IIL, IIP and SNe  Ia, Ib, Ic.
As already mentioned before, SNe II,  which
are believed to be the end state of stars more massive than 
$10 M_{\odot}$ exploding after a Fe core is formed (core-collapse SNe),
produce mainly $\alpha$-elements (O, Ne, Mg, Si, S, Ca) plus
some Fe. The amount of Fe produced by type II SNe is one of the most uncertain
quantities since it depends upon the 
so-called mass cut (how much Fe remains in the collapsing core and how 
much is ejected)
and on explosive nucleosynthesis. 

Type Ia SNe are believed to originate from the C-deflagration of a 
WD reaching the Chandrasekhar mass (1.44 $M_{\odot}$)
after accretion of material from a young companion in a close binary system.
C-deflagration occurs as a consequence of such accretion and an 
explosion ensues destroying the whole star. No remnant is then left behind.  
They produce a large amount 
($\sim 0.6-0.7 M_{\odot}$)
of $^{56}Ni$ (Fe) plus traces of C to Si elements. C-deflagration is 
the explosive burning which best reproduces the observed abundance pattern 
in Type Ia SN remnants. The best model for this kind of explosion 
is model W7 by Nomoto, Thielemann \& Yokoi (1984).
The amount of Fe produced by the other Type I SNe is smaller than 
that produced by the Type Ia ones at least a factor of two or more. 
Therefore,
Type Ia SNe should be considered as the responsible for the Fe 
enrichment in the universe.

\subsection{Element production}
Here is a summary of element production:

\begin{itemize}

\item Big Bang $\rightarrow$ light elements H, D, $^{3}He$, $^{4}He$,
$^{7}Li$. Deuterium is only destroyed inside stars to form $^{3}He$.
$^{3}He$ is also mainly destroyed. The only stars producing some 
$^{3}He$ are those with masses  $< 2.5 M_{\odot}$.
Recent prescriptions for the  yields of $^{3}He$ are
from Forestini \& Charbonnel (1997) and Sackmann \& Boothroyd (1999).
Lithium: $^{7}Li$ is produced during the Big Bang but also 
in stars: massive AGB stars, SNe II, carbon-stars and  novae.
Some $^{7}Li$ should also be produced in spallation processes by 
galactic cosmic rays (see Romano et al. 2001).
\item Spallation Processes $\rightarrow$ $^{6}Li$, Be and B.

\item   Type II SNe $\rightarrow$ $\alpha$-elements 
(O, Ne, Mg, Si, S, Ca), some Fe,  s-process elements ($A< 90$)
and r-process elements. 
Yields are from Woosley \& Weaver (1995) and Thielemann et al. (1996).

\item Type Ia SNe $\rightarrow$ Fe-peak elements. Yields are from 
Nomoto et al. (1984) and Nomoto et al. (1997).

\item Low and intermediate mass stars $\rightarrow$ $^{4}He$,
C, N, s-process ($A>90$)
elements. Yields are from Renzini \& Voli (1981), Marigo et al. (1996),
van den Hoek \& Groenewegen (1997) and Gallino et al. (1998).

\end{itemize}

\subsection{Stellar yields}
In order to include the results from nucleosynthesis into chemical 
evolution models we need to define the stellar yields.
The stellar yield of an element $i$ is defined as the mass fraction 
of a star of mass $m$ which has been newly created as species $i$ and ejected:

\begin{equation}
p_{im}=({M_{ej} \over m})_{i}
\end{equation}

In order to compute $p_{im}$ we need to know some 
fundamental quantities from stellar evolution and nucleosynthesis:
\begin{itemize}
\item
$M_{\alpha}$ is the mass of the He-core 
(where H is turned into He)

\item $M_{CO}$ is the mass of the C-O core (where the He is turned 
into heaviers)

\item $M_{rem}$ is the mass of the remnant (WD, neutron star, 
black hole)

\end{itemize}

These  masses are related to each other  by the following relations:
$$
M_{\alpha}-M_{He}=M_{CO}
$$
where $M_{He}$ is the newly formed and ejected $^{4}He$ and:
$$
M_{CO}-M_{r}=M_{C}+M_{O}+M_{heaviers}
$$
The values of these different quantities are given by stellar evolution 
and nucleosynthesis calculations.

\section{Modelling chemical evolution}
\subsection{Analytical models}
The simplest model of chemical evolution is the 
{\it Simple Model} for
the chemical evolution of the solar neighbourhood.
The basic assumptions of the Simple Model are:\par
- the system is one-zone and closed, namely there are no inflows 
or outflows,\par
- the initial gas is primordial (no metals),\par
- $\varphi(m)$ is constant in time,\par
- the gas is well mixed at any time.\par

In the following we will adopt the
formalism of Tinsley (1980) and define:
\begin{equation}
\mu={M_{gas} \over M_{tot}} 
\end{equation}
as the fractional mass of gas, with:

\begin{equation}
M_{tot}=M_{*}+M_{gas} 
\end{equation}
where $M_*$ is the mass in stars (dead and alive). 
Possible non-baryonic dark matter is not considered.\par
The mass of stars can be expressed as:
\begin{equation}
M_*=(1- \mu) M_{tot}. 
\end{equation}
The abundance by mass of an element $i$ is defined by:
\begin{equation}
X_i={M_{i} \over M_{gas}} 
\end{equation}
where $M_{i}$ is the mass in the form of the specific element $i$.
It is well known that the 
abundances must satisfy the condition,
$\sum_{i}{X_i(t)}=1$,
where the summation is over all the chemical elements.

The initial conditions are:
\begin{equation}
M_{gas}(0)=M_{tot} \,\,\,\,\,;
X_i(0)=0, 
\end{equation}
where $i$ refers to metals.
The equation for the evolution of the gas in the
system can be written as:
\begin{equation}
{dM_{gas} \over dt}= -\psi(t)\,\,+ E(t) 
\end{equation}
where $E(t)$ is the
rate at which dying stars restore both the enriched and unenriched
material into the ISM.
$E(t)$ can be written as:
\begin{equation}
E(t)= \int^{\infty}_{m(t)}{(m-M_{rem}) \psi(t-\tau_m) \varphi(m) dm} 
\end{equation}
where $m-M_{rem}$
is the total mass ejected from a star of mass $m$, and $\tau_m$ is 
the lifetime of a star of mass $m$.
When $E(t)$ is substituted into the gas equation one obtains
an integer-differential
equation which can be solved analytically only by assuming
Instantaneous Recycling Approximation (I.R.A.).
In this approximation one assumes that all stars less massive than 
$1 M_{\odot}$ live forever whereas all stars more massive than 
$1 M_{\odot}$ die instantaneously. In other words, I.R.A. allows us 
to neglect the stellar lifetimes and solve analytically equation (4.31). 
Under I.R.A., 
we can define the returned fraction:
\begin{equation}
R=\int^{\infty}_{1}{(m-M_{rem}) \varphi(m) dm} 
\end{equation}
which is called fraction because is divided by $\int^{\infty}_{1}{m \varphi(m) dm}=1$, which depends on the normalization of the IMF. 
We also define the yield per stellar generation as:
\begin{equation}
y_{i}={1 \over 1-R} \int^{\infty}_{1}{m p_{im} \varphi(m) dm} 
\end{equation}
where $p_{im}$ is the stellar yield previously defined.
Then, by substituting R and $y_i$ into equation (4.32) we obtain:
\begin{equation}
E(t)= \psi(t) R 
\end{equation}
and:
\begin{equation}
{dM_{gas} \over dt}= -\psi(t)(1-R) 
\end{equation}

The equation for the evolution of the chemical abundances can be written as:
\begin{equation}
{d(X_iM_{gas}) \over dt}=-X_i\psi(t) \,\,+ E_{i}(t) 
\end{equation}
where:
\begin{equation}
E_{i}(t)=\int^{\infty}_{m(t)}[(m-M_{rem})X_i(t-\tau_m)+
mp_{im}] \cdot
\psi(t-\tau_m) \varphi(m)dm 
\end{equation}
contains both the unprocessed and the newly produced element $i$.
It is worth noting that this equation is valid for
metals and not for elements which are 
wholly or partly destroyed in stars.
Under the assumption of I.R.A. the above eq. becomes:
\begin{equation}
E_{i}(t)=\psi(t)RX_i(t) + y_{i}(1-R) \psi(t)
\end{equation}
When substituted in (4.37) the equation
can be solved analytically with the previous initial conditions
and the solution is:
\begin{equation}
X_i= y_{i} ln({ 1 \over \mu}) 
\end{equation}
the famous solution for the {Simple Model}.
The yield which appears in the above solution
is known as {\it effective yield}, 
simply defined as the yield $y_{i_{eff}}$ that would be deduced 
if the system were assumed to be described by the Simple Model:

\begin{equation}
y_{i_{eff}}={X_i \over ln(1/\mu)} 
\end{equation}
The meaning of the effective yield can be understood with the 
following example:
if $y_{i_{eff}} > y_{i}$(true yield) 
then the actual system has attained a higher
abundance for the element $i$ at a given gas fraction $\mu$.

\subsection{Failure of the Simple Model}
The Simple Model predicts too many stars with metallicity
lower than [Fe/H]= -1.0 dex relative to observations.
This is known as ``G-DWARF PROBLEM''.
However, the G-dwarf is no more a problem 
since several solutions  
have been suggested.

Possible solutions to the G-dwarf problem include:
\par
-Slow formation of the solar vicinity by gas infall\par
-Variable IMF\par
-Pre-enriched gas \par

Generally, the slow infalling gas is preferred since it is the most 
realistic suggestion and produces results in very good agreement with the 
observations, as we will see in the following.

\subsection{Analytical models with gas flows}

The equation for the evolution of abundances in presence of gas flows 
transforms into:

\begin{equation}
{d(X_iM_{gas}) \over dt}=-X_i(t)\psi(t) + E_{i}(t)
+  X_{A_{i}}A(t) 
- X_i(t)W(t) 
\end{equation}
where
$A(t)$ is the accretion rate of matter
with abundance of the element $i$ $X_{A_{i}}$ and
$W(t)$ 
is the rate of loss of material from the system.
The case $A(t)=W(t)=0$ obviously corresponds to the Simple Model.
The case $A(t)=0$, $W(t) \ne 0$ corresponds to the outflow model.
The easiest way of defining $W(t)$ in order to solve the equation 
analytically is to assume:
\begin{equation}
W(t)=\lambda(1-R) \psi(t) 
\end{equation}
where $\lambda \ge 0$ is the wind parameter.
The analytical solution for the equation of metals,
which can be integrated between 0 e $X_i(t)$ and 
between $M_{gas}(0)=M_{tot}$ and  $M_{gas}(t)$, is:
\begin{equation}
X_i={y_i  \over (1+ \lambda)} ln [(1+ \lambda)\mu^{-1}- \lambda] 
\end{equation}
For $\lambda=0$ we recover the solution of the Simple Model.

The case of 
$A(t) \ne 0$ and $W(t)=0$ corresponds to the accretion model.
The easiest way to choose the
accretion rate is:
\begin{equation}
A(t)=\Lambda (1-R) \psi(t) 
\end{equation}
with $\Lambda$ being a positive constant different from zero.
The solution of the equation of metals for
a primordial infalling material ($X_{A_{i}}=0$) and $\Lambda \ne 1$  
is :
\begin{equation}
X_i= {y_i \over \Lambda}[1-(\Lambda-(\Lambda-1)\mu^{-1})^{-\Lambda/(1-\Lambda)}]
\end{equation}
as shown by
Matteucci and Chiosi (1983).
If $\Lambda=1$ ( extreme infall model) the solution is:
\begin{equation}
X_i=y_i[1-e^{-(\mu^{-1}-1)}] 
\end{equation}
where the quantity
$\mu^{-1} -1$ represents the ratio between the accreted mass 
and the initial mass. 

\section{Equations with Type Ia and II SNe}
In general, if one wants to compute in detail the evolution of the 
abundances of elements produced and restored into the ISM on long timescales, 
the I.R.A. approximation is a bad approximation. Therefore, it is 
necessary to 
consider the stellar lifetimes in the chemical evolution equations and 
solve them with numerical methods.

If $G_i$ is the mass fraction of gas in the form of an element $i$, 
we can write:
\begin{eqnarray}
 & & \dot G_i(t)  =  -\psi(t)X_i(t)\nonumber\\
& & + \int_{M_{L}}^{M_{Bm}}\psi(t-\tau_m)
Q_{mi}(t-\tau_m)\phi(m)dm \nonumber\\ 
& & + A\int_{M_{Bm}}^{M_{BM}}
\phi(m)\nonumber \\
& & \cdot[\int_{\mu_{min}}
^{0.5}f(\mu)\psi(t-\tau_{m2}) 
Q_{mi}(t-\tau_{m2})d\mu]dm\nonumber \\ 
& & + B\int_{M_{Bm}}^
{M_{BM}}\psi(t-\tau_{m})Q_{mi}(t-\tau_m)\phi(m)dm\nonumber \\
& & + \int_{M_{BM}}^{M_U}\psi(t-\tau_m)Q_{mi}(t-\tau_m) 
\phi(m)dm \nonumber\\ 
& & + X_{A_{i}} A(t) - X_{i} W(t)
\end{eqnarray}
where B=1-A.The parameter A represents the unknown fraction of binary stars
giving rise to type Ia SNe and is fixed by reproducing the observed 
present time SN Ia rate.Generally, values in the range $A=0.05-0.09$
(according to the IMF)
reproduce well the observed SN Ia rate in the Galaxy.
The chemical abundances are defined as:
$X_{i}={G_i \over G}$, where $G=\mu={M_{gas} \over M_{tot}}= 
{\sigma_{gas} \over \sigma_{tot}}$.
The total mass $M_{tot}$ (or surface mass density $\sigma_{tot}$) 
refers to the mass of 
stars (dead and alive) plus gas at the present time.

$W_{i}(t)=X_{i} W(t)$  
is the galactic wind rate for the element $i$, whereas  
$A_{i}(t)=A(t) X_{A_{i}}$, with $A(t)$ being the accretion rate  
for the element $i$.

$M_{Bm}$ is the total minimum and $M_{BM}$ the total maximum  
mass allowed for binary systems giving rise to Type Ia SNe 
(Matteucci \& Greggio 1986). 
For the model of a C-O WD plus a red-giant companion for the
progenitors of Type Ia SNe, 
$M_{BM} \le
16 M_{\odot}$. 
$M_{Bm}$ is more uncertain and often has been taken to be 
$3M_{\odot}$ to ensure that the primary star 
(the initially more massive in the binary system) 
would be massive 
enough to guarantee that after accretion from the companion the C-O 
white dwarf eventually reaches the Chandrasekhar mass and ignites carbon.
This formulation of the SN Ia rate was originally proposed by 
Greggio and Renzini (1983). 
Greggio (1996) presented revised criteria for the choice of $M_{Bm}$.
In particular, the suggested condition for the explosion of the system is:
\begin{equation}
M_{WD} +\epsilon M_{2,e} \ge M_{Ch} 
\end{equation}
(where $M_{2,e}$ is 
the envelope mass of the evolving secondary, $M_{WD}$ is the 
mass of the white dwarf and $M_{Ch}$
is the Chandrasekhar mass). 

For models involving Sub-Chandrasekhar 
white dwarf masses, which have been suggested to explain subluminous 
Type Ia SNe Greggio obtains:
$
M_{WD} \ge 0.6
M_{\odot}
$
and 
$
\epsilon M_{2,e} \ge 0.15 
$

The masses $M_L=0.8$ and $M_{U}=100M_{\odot}$ define the lowest and 
the highest mass, respectively, contributing to the chemical enrichment.
The function $\tau_m(m)$ 
describes the stellar lifetimes.
The quantity $Q_{mi}(t-\tau_m)$
contains all
the information about stellar 
nucleosynthesis for elements either produced or destroyed inside 
stars or both, and is defined as in 
Talbot and Arnett (1973).

\subsection{Type Ia SN rates}
The single degenerate (SD) scenario is
based on the
original suggestion of Whelan and Iben (1973), namely C-deflagration in
a WD reaching the Chandrasekhar mass after accreting material from a
star which becomes red giant and fills its Roche lobe.
An alternative to the SD scenario is represented by 
the double degenerate (DD) scenario, where
the merging of two C-O WDs, due to gravitational wave radiation,
creates an object exceeding the Chandrasekhar mass and 
exploding by C-deflagration (Iben
and Tutukov 1984).
Negative results from observational searches for very close binary
systems made of massive enough WDs (Bragaglia et
al. 1990) has made the DD scenario less attractive and people 
to concentrate more on the SD scenario.

A recent model has been suggested by Hachisu et al.(1996; 1999) and 
is based on the
classical scenario of Whelan and Iben (1973) 
but with a
metallicity effect. It predicts that no Type Ia systems can form 
for [Fe/H]$< -1.0$. This model seems to have some difficulty in explaining
the low [$\alpha$/Fe] ratios observed in Damped Lyman- $\alpha$ systems 
(DLAs) which show that even at low metallicities is present the effect of 
Type Ia SNe.
Recently, Matteucci \& Recchi (2001) showed that there are some problems 
with this scenario also in explaining the observed [O/Fe] vs. [Fe/H] relation
in the solar neighbourhood and concluded that the best model is still 
the SD one in the formulation proposed by Greggio \& Renzini (1983).
They also showed that  
the typical timescale for Type Ia SN enrichment, defined
as the time $t_{SNIa}$ when the SN rate reaches the maximum,
varies strongly from galaxy to galaxy and that it is not correct to adopt 
a universal $t_{SNIa}$=1 Gyr, as often quoted in the literature.
Their results indicate that 
for an elliptical galaxy with high SFR, 
$t_{SNIa}= 0.3-0.5$ Gyr,
for a spiral Galaxy like the Milky Way,
$t_{SNIa}=4-5$ Gyr and
for an irregular galaxy with a continuous but very low SFR, 
$t_{SNIa} > 5$ Gyr.
This fact has very important consequences on the chemical evolution of 
galaxies of different morphological types, as we will see in the following.

\section{The formation and evolution of the Milky Way}
The first application of models of chemical evolution is the Milky Way (MW)
Galaxy
for which we have most of the available data.
Before describing the chemical evolution of the Galaxy I recall some of the
ideas proposed for the formation of the MW.
Eggen, Lynden-Bell \& Sandage (1962) suggested a rapid collapse for 
the formation of the Galaxy lasting $\sim 3 \cdot 10^{8}$ years
implying that no spread in the age of Globular Clusters should be observed.
They based their suggestion on the finding that halo stars have high 
radial velocities whitnessing the initial fast collapse.
Later on, Searle \& Zinn (1978) proposed a central collapse
but also
that the outer halo formed by merging of large fragments taking place 
over a considerable timescale $> 1$ Gyr.
More recently,
Berman \& Suchov (1991) proposed the {\it hot Galaxy picture},
an initial strong burst of star formation which inhibited  
further star formation  for a few gigayears, 
while a strong galactic wind was created. 
Subsequently, the remainder of the proto-Galaxy, contracted and cooled to 
form the major stellar components 
observed today.
At the present time, the most popular idea on the formation of the Galaxy 
is that the inner halo formed rather quickly on a 
time scale
of 0.5-1 Gyr, whereas the outer halo formed more slowly by mergers of 
fragments or accretion from satellites of the MW (see Matteucci 2001).

\subsection{Models for the Milky Way}
\begin{itemize}

\item{SERIAL FORMATION APPROACH:}
halo, thick and thin disk formed in sequence, as a continuous process
(e.g. Matteucci  \& Fran\c cois 1989). 

\item{PARALLEL FORMATION APPROACH:}
the various Galactic components start forming at the same time and 
from the same gas but evolve at different rates
(e.g. Pardi, Ferrini \& Matteucci 1995). At variance with 
the previous scenario
it predicts overlapping of stars belonging to the different components.

\item{TWO-INFALL APPROACH:}
the evolution of the halo and disk are totally independent and they 
form out of two separate infall episodes (overlapping in metallicity 
is also predicted) 
(e.g. Chiappini, Matteucci \& Gratton, 1997; Chang et al. 1999, Alib\`es 
et al. 2001).

\item{STOCHASTIC APPROACH:}
in the early halo phases, mixing was not efficient, thus
pollution from single SNe (Tsujimoto et al. 1999; Argast et al. 2000; 
Oey 2000) can be seen in very metal poor stars. This approach predicts a 
large spread in the abundance ratios at very low [Fe/H], even larger 
than observed. 

\end{itemize}
\subsection{The two-infall model}
Here I describe in more detail the two-infall 
approach (Chiappini et al. 1997) since 
it gives the best agreement with observations for the halo and disk.

The basic equations for the evolution of the MW in this case contain
only an accretion term (no outflow):
$${dG_i(r,t) \over dt} = -X_i \psi(r,t) + $$ \par
$$(1 - \alpha) \int^{M_{Bm}}_{M_L}{
\psi(r, t- \tau_m) Q_{m_i}(t- \tau_m) \varphi(m) dm} + $$\par 
$$ \alpha n \int^{M_{Bm}}_{M_L}{\psi(r, t- \tau_m - \Delta t) 
Q_{m_i}(t- \tau_m) 
\varphi(m) dm}+$$ \par
$$ A\int^{M_Bm}_{M_{BM}}{SNIa} +$$\par
$$(1-A) \int^{M_Bm}_{M_{BM}}{SNeII} + \int^{M_U}_{M_{BM}}{SNeII}+ $$ \par
$$({d G_i \over dt})_{inf} $$

These equations are the same as (5.48) plus a term which contains 
the contribution from novae,
which are perhaps important producers of $^{7}Li$, $^{15}N$ 
and $^{13}C$
(D'Antona \& Matteucci, 1991;Romano et al. 2001). 
The nova contribution contains the parameter
{$\alpha$=0.0155 which represents the fraction of WDs
which are in binary systems giving rise to nova events. The value of
$\alpha$ is chosen to reproduce the present time nova rate 
$R_{novae} \sim 26 yr^{-1}$},
after assuming that each nova has $\sim 10^{4}$ outbursts 
all over its lifetime.
The quantity
$\Delta t \sim 1$ Gyr is the assumed time-delay for the starting 
of nova activity since the formation of the binary system.

The infall rate term is defined as:

\begin{equation}
A_i(t)={dG_i(r,t) \over dt}={A(r)(X_{A})_{i} e^{-t/\tau_H} \over \sigma_{tot}(r,t_G)}+
{B(r)(X_{A})_{i} e^{-(t-t_{max})/ \tau_D(r)} \over \sigma_{tot}(r,t_G)}
\end{equation}

where $\tau_H$ is the timescale for the inner halo formation (0.5-1 Gyr)
and  $\tau_D(r)$ is the thin disk timescale varying with galactocentric 
distance (inside-out formation):

\begin{equation}
\tau_D(r)=0.875r-0.75
\end{equation}

The SFR is given by:
\begin{equation}
\psi(t)=\nu \sigma_{tot}(r,t)^{k_1} \sigma_{gas}(t)^{k_2}
\end{equation}

with $k_1=0.5$ and $k_2=1.5$.
The behaviour of this SFR for the halo-thick disk and the thin-disk phase, 
respectively, is show in Figure 1.
A threshold density ($\sigma_{th}=7 M_{\odot} pc^{-2}$) 
in the SFR is assumed. 

\begin{figure}[t!]
\resizebox{\hsize}{!}{\includegraphics{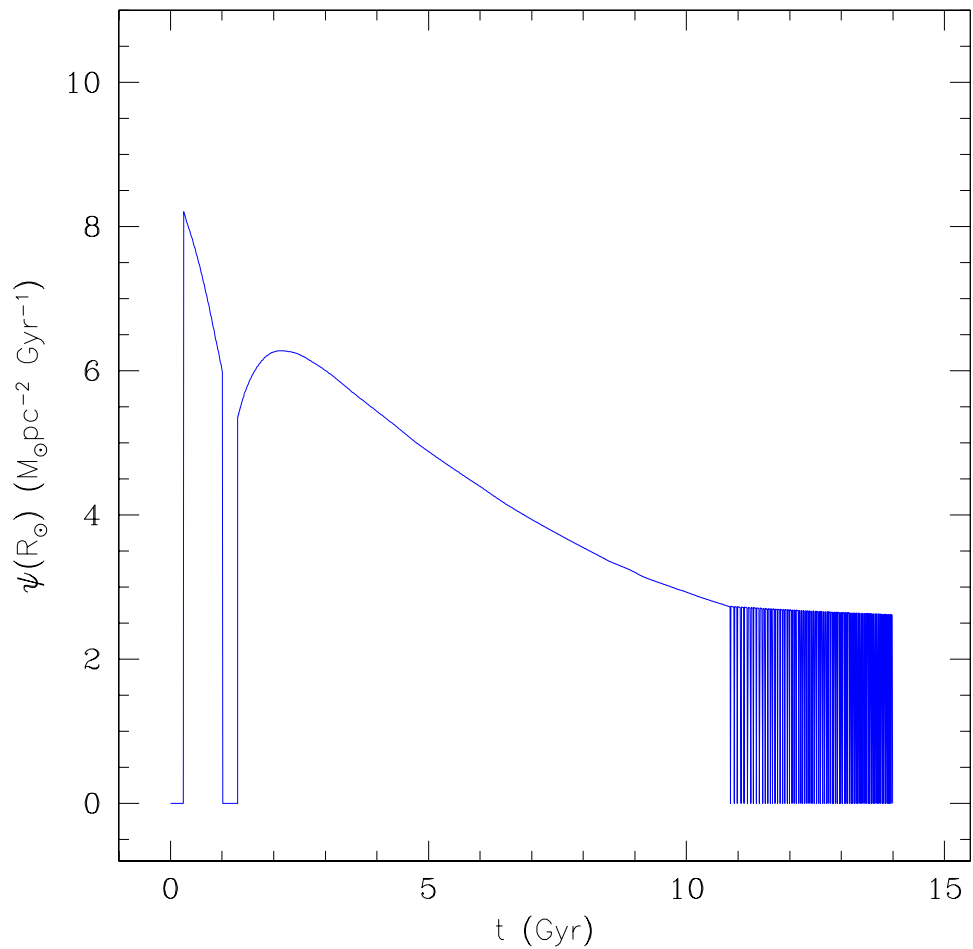}}
\hfill
\parbox[b]{\hsize}{
\label{fig1} 
\caption{Predicted star formation rate in the halo-thick disk phase and 
in the thin disk phase during the evolution of the Milky Way.
Notice that because of the existence of a threshold in the gas density 
the SFR halts in between the two major episodes of infall and
oscillates in the last phases of the evolution of the disk. 
The models are from Chiappini et al. (2001).}} 

\end{figure}

\subsection{Applications to the Local Disk}
\begin{itemize}
\item {\it The G-dwarf metallicity distribution} is shown in Figure 2
where the data are compared with the predictions of the two infall model 
assuming a time scale for the formation of the local disk of 8 Gyr.
(Chiappini et al. 1997;
Boissier and Prantzos 1999; Chang et al. 1999; Chiappini et al. 2001).

\begin{figure}[t!]
\resizebox{\hsize}{!}{\includegraphics{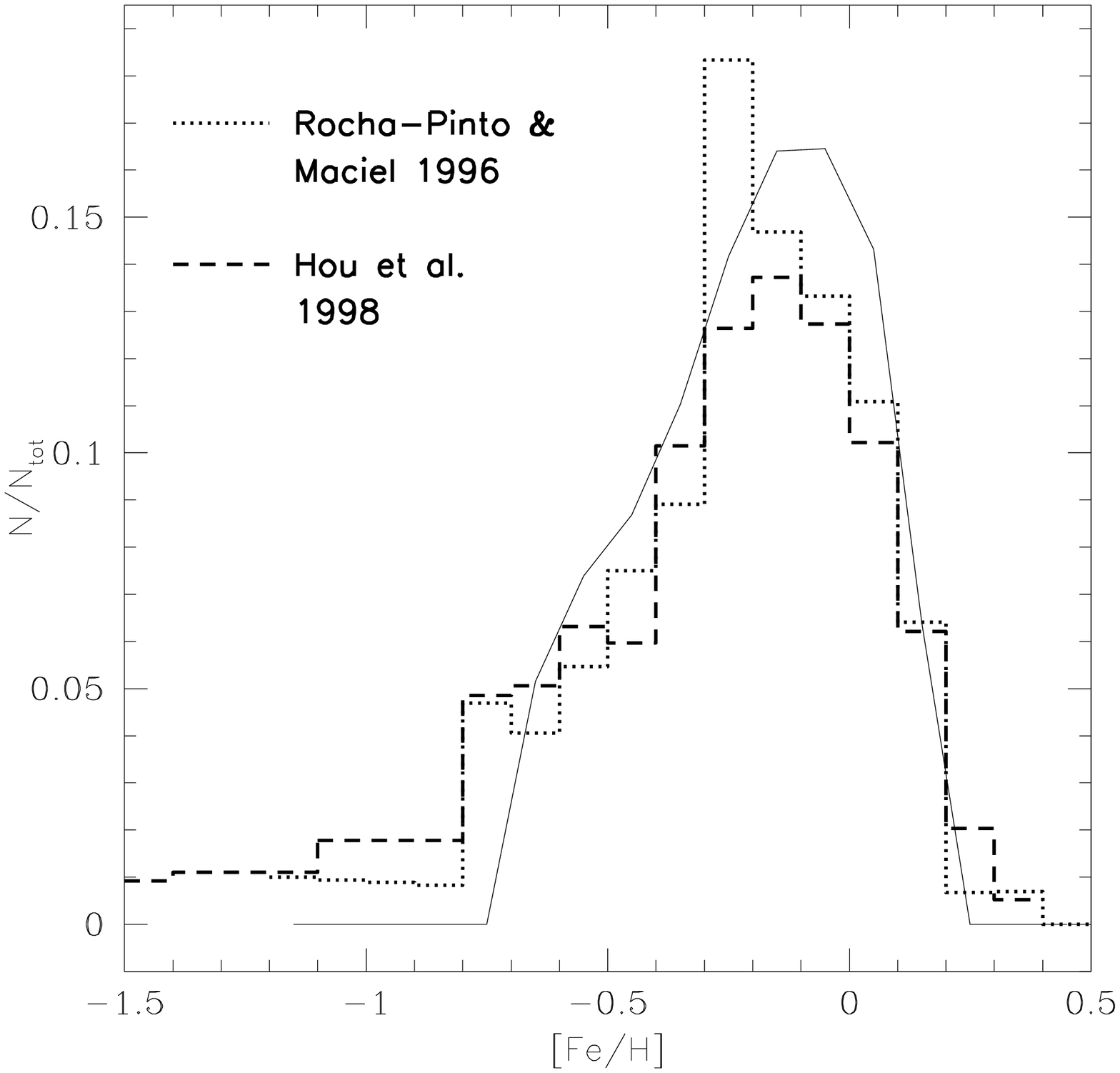}}
\hfill
\parbox[b]{\hsize}{
\label{fig2} 
\caption{Observed and predicted G-dwarf metallicity distribution.
The data are from Rocha-Pinto and Maciel (1996) and Hou et al. (1998).
The model predictions (continuous line) are from Chiappini et al. (2001).
The model assumes that the timescale for the disk formation in the 
solar neighbourhood is 8 Gyr.}}
\end{figure}

\item {\it The relative abundance ratios as functions of the relative 
metallicity (relative to the Sun) [X/Fe] vs. [Fe/H]}, they are
interpreted as
due to the time-delay between Type Ia and II SNe. In fact, for the 
$\alpha$-elements the slowly declining [$\alpha$/Fe] 
ratio at low [
Fe/H] (halo phase) is normally interpreted as due to the pollution from 
massive stars, whereas the abrupt change in slope occurring at 
$\sim [Fe/H] =-1.0$ dex is due to the Type Ia SNe restoring the bulk of iron. 
From the [$\alpha$/Fe] vs. [Fe/H] diagram one can infer the timescale 
for the formation of the halo
($\tau_h \sim$ 1.5-2.0 Gyr, Matteucci and Fran\c cois, 1989; 
Chiappini et al. 1997), just by means of the age-[Fe/H] relationship, 
which indicates the time at which the metallicity of the 
turning point is reached.
In Figure 3 we do not show the usual plot but [Fe/O] vs. [O/H] since 
in this plot are evident some features which do not appear in the 
classical diagram.
In particular, the data show evidence for 
a gap around [O/H]= -0.3 dex, corresponding to [Fe/H] 
$\sim$ -1.0 dex (Gratton et al. 2000).
This gap is well reproduced by the model which predicts a hiatus 
(no more than 1 Gyr) 
in the SFR between the end of the thick-disk phase and the beginning of
the formation of 
the thin disk. This hiatus produces, in fact, a situation 
where O is no longer produced whereas Fe is produced, and this is revealed 
by the increase of [Fe/$\alpha$] at constant [$\alpha$/H]. 
This effect has been observed also for [Fe/Mg] vs. [Mg/H] by 
Furhmann (1998). The model shown in Figure 3 is a two-infall model 
with a threshold density for the star formation and is just the 
existence of such a threshold which produces the hiatus in the SFR, 
evident in Figure 1.

\end{itemize}

\begin{figure}[t!]
\resizebox{\hsize}{!}{\includegraphics{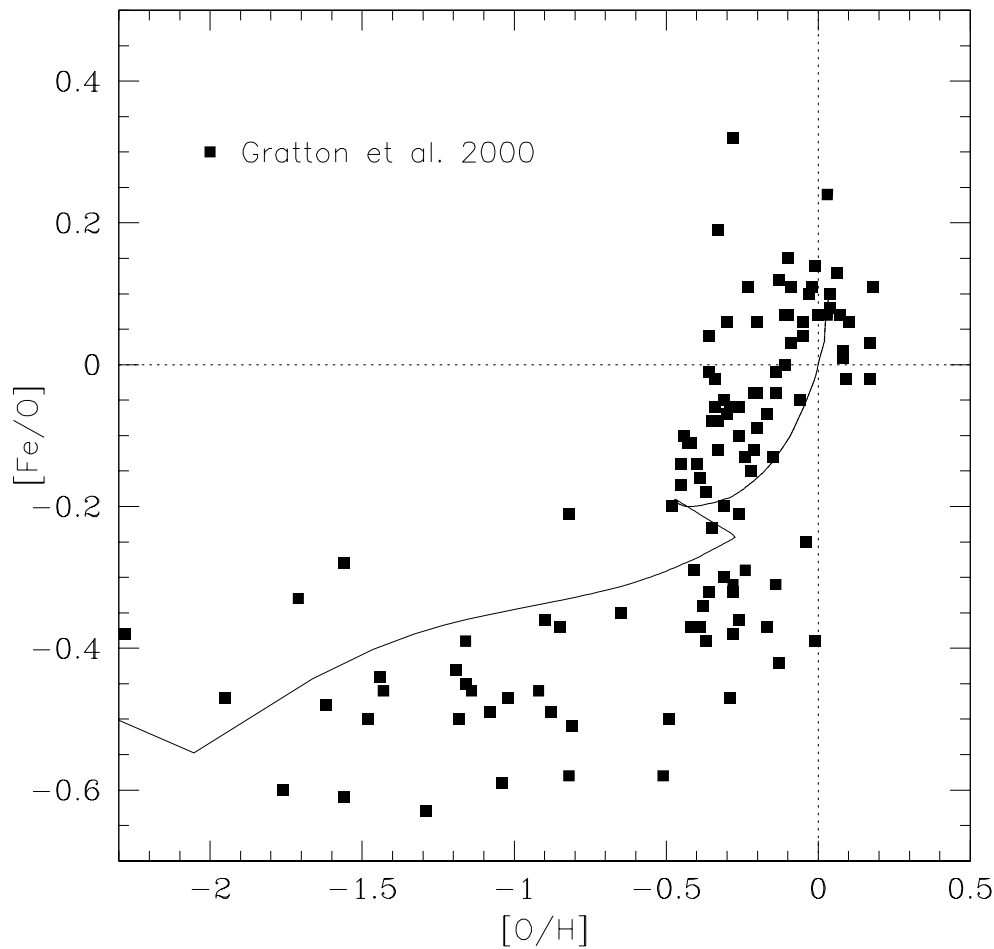}}
\hfill
\parbox[b]{\hsize}{
\label{fig3} 
\caption{The [Fe/O] vs. [O/H] relation in the solar neighbourhood. 
The data are from Gratton et al.(2000) and the model (continuous line) 
from Chiappini
et al.(2001). In this figure is evident the existence of a gap in the data 
at around [O/H]=-0.3 dex, corresponding to
[Fe/H] $\sim$ -1.0 dex, which is also predicted by the model.}}
\end{figure}

\subsection{Applications to the whole disk }

\begin{itemize}
\item 
{\it Abundance Gradients} are known to exist along the Galactic disk from
data from various sources (HII regions, PNe, B stars). These data
suggest that the gradient for oxygen is $\sim -0.07$ dex/kpc in the 
galoctocentric distance range 4-14 kpc. 
It is not yet clear if the slope is unique
or if there is a change in slope as a function of the galactocentric
distance. Similar gradients are found for N and Fe 
(see Matteucci 2001 and references therein).

\item {\it Gas Distribution}. HI is roughly constant 
over a range of 4-10 kpc along the Galactic disk
while $H_2$ follows the light distribution. No models
can explain the two distributions. The total gas increases 
towards the center with a peak at 4-6 kpc.

\item {\it The SFR Distribution} is obtained 
from various tracers (Lyman-$\alpha$ continuum, pulsars, 
SN remnants, molecular clouds) and shows that the SFR increases with 
decreasing galactocentric distance reaching a peak 
at 4-6 kpc in correspondence of the gas peak.

\end{itemize}

In order to fit gradients, SFR and gas one has to assume  
that the disk formed inside-out, in agreement with 
a previous suggestion by Larson (1976) and that
the SFR should be a strongly varying function of the 
galactocentric distance.
In Figure 4 is shown a comparison between an inside-out model and the
abundance gradient, the gas, star and SFR distributions.

\begin{figure}[t!]
\resizebox{\hsize}{!}{\includegraphics{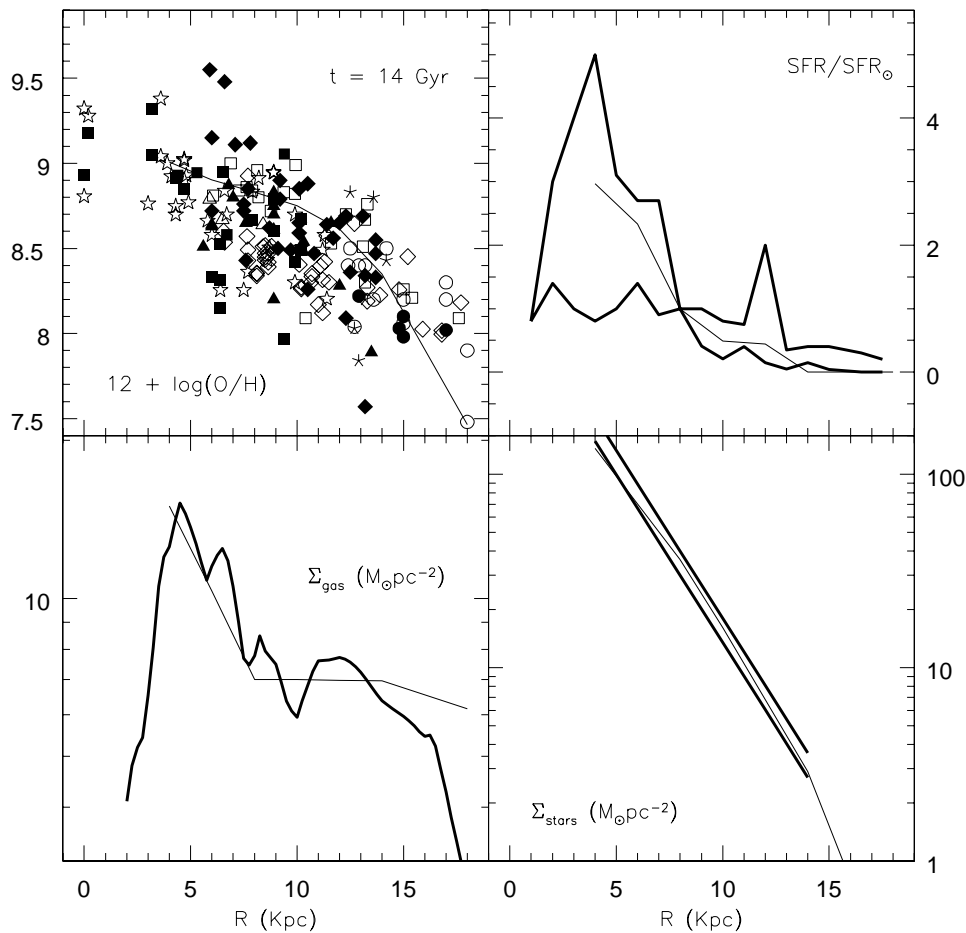}}
\hfill
\parbox[b]{\hsize}{
\label{fig4} 
\caption{Comparison between data and model predictions from Chiappini 
et al. (2001) (continuous lines).
In the first panel from the top we show the oxygen abundance gradient. 
In the second panel the variation of the $SFR/SFR_{\odot}$,
in the third panel the gas distribution and in the fourth panel 
the distribution of stars along the Galactic disk.}}
\end{figure}

\subsection{The Role of Radial Flows in the evolution of the Galactic Disk}
The gas infalling onto the disk can induce radial inflows by transferring angular momentum to the gas in the disk. Angular momentum 
transfer can be due to the gas viscosity in the disk and
induce inflows in the inner parts of the disk and outflows in the outer parts.
All viscous models suggest that metallicity gradients can be 
steepened by radial (in)flows, especially if an outer star formation 
cut-off is assumed
(Clarke 1989;
Yoshii \& Sommer-Larsen 1989).
All models agree that the velocity of radial (in)flows 
should be low 
($v< 2$ km/sec). Observationally is not clear if these radial 
flows exist.
Edmunds \& Greenhow (1995), by means of analytical models of galactic 
chemical evolution concluded that
there is no simple {\it one-way} effect of radial flows on abundance 
gradients.
Portinari \& Chiosi (2000) suggested that radial inflows 
may represent 
a possible explanation of the peak of the gas at 4-6 kpc.
In our opinion radial flows, if they exist, are never the main cause 
for the formation of abundance gradients.

\subsection{The Role of the IMF in the evolution of the Galactic Disk}

Abundance gradients, in principle can be obtained by a variation of the IMF along the disk. One can assume either 
that more massive stars form in external regions or the contrary,
that more low mass stars form in external regions. 
It has been shown convincingly that neither of the two options work
(Carigi 1996; Chiappini et al. 2000), and that there is no evidence 
in favor of such variations along the Galactic disk. Therefore, 
we can exclude the variation of the IMF as the main 
cause of abundance gradients
and conclude that the best agreement with the observed properties 
of the Galactic disk is obtained by assuming a constant IMF.

\subsection{Scenarios for Bulge Formation}
Various scenarios have been proposed insofar for the formation of the 
Galactic bulge but only one seems to reproduce the observed abundance pattern.
Here I recall the main scenarios:
\begin{itemize}
\item Accretion of extant stellar systems which eventually settle in the center of the Galaxy.

\item Accumulation of gas at the center of the Galaxy 
and subsequent evolution with either fast or slow star formation.

\item Accumulation either rapid or slow 
of metal enriched gas from the halo or thick disk
in the Galaxy center.

\item Formation occurs  out of inflow of metal enriched 
gas from the thin-disk.
\end{itemize}
The metallicity distribution of stars in the bulge as well as
the [$\alpha$/Fe] vs. [Fe/H] relations help 
in selecting the most probable scenario and suggest that a fast 
accumulation of gas in the Galactic center accompanied by fast 
star formation
is the best scenario.
In this framework, the bulge must have formed contemporarily to the 
inner halo on a similar timescale.
In Figures 5 and 6 we show some model predictions compared with the 
metallicity distribution of bulge stars and the predicted [$\alpha$/Fe] 
vs. [Fe/H] ratios for the bulge together with the predictions for the 
solar neighbourhood.
The model for the bulge presented in the two figures
(Matteucci, Romano \& Molaro, 1999) assumes a much stronger star 
formation rate than in the solar neighbourhood (by a factor of 10) 
with the same nucleosynthesis prescriptions and a timescale for bulge 
formation of 0.5 Gyr as opposed to 8 Gyr in the solar vicinity.
In Figure 5 it is evident that the best model to reproduce the stellar 
metallicity distribution requires a Salpeter (1955) IMF which is 
sligthly flatter than that used for the solar neighbourhood (Scalo, 1986).
The predicted [$\alpha$/Fe] ratios in Figure 6 indicate that the bulge 
stars should show overabundances of $\alpha$-elements for most of the [Fe/H] 
range, as it seems also suggested by observations (Mc William \& Rich, 1994; 
Barbuy, 1999). This is a consequence of the time-delay between 
Type Ia and II SNe, coupled with a very fast evolution in the bulge 
as compared to the solar vicinity. In fact, in this case high values 
of [Fe/H] are reached in the 
gas before a substantial number of SNe Ia has the time to restore the bulk 
of iron.
The contrary occurs in a system with lower star formation than in the 
solar neighbourhood, such as the external regions of the disk and 
irregular galaxies.
In these systems we expect that the overabundance of $\alpha$-elements 
relative to Fe is maintained only for a short interval of [Fe/H] 
(see Pagel, 1997: Matteucci 2001).

\begin{figure}[t!]
\resizebox{\hsize}{!}{\includegraphics{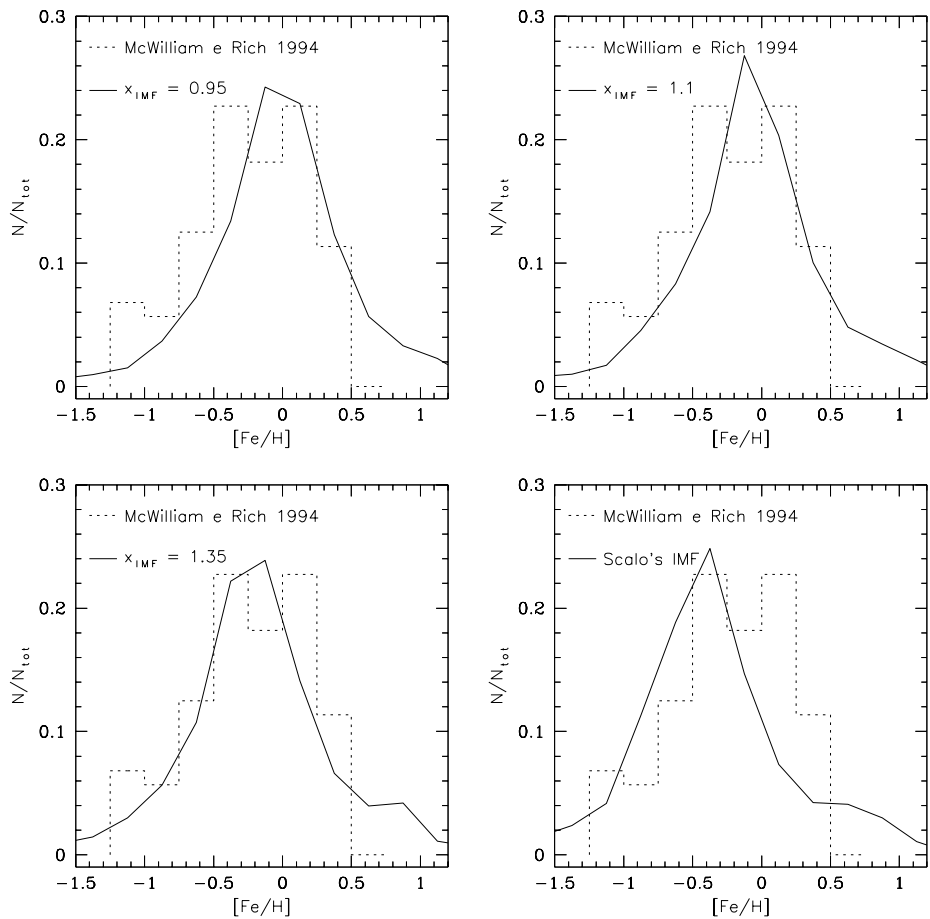}}
\hfill
\parbox[b]{\hsize}{
\label{fig5} 
\caption{Comparison between data (dotted lines) and model predictions 
for the metallicity distribution of bulge stars. The data are from 
McWilliam \& Rich (1994) and the models (continuous lines) from 
Matteucci et al. (1999).
Each panel corresponds to a model with a different IMF slope, as indicated.}} 
\end{figure}

\begin{figure}[t!]
\resizebox{\hsize}{!}{\includegraphics{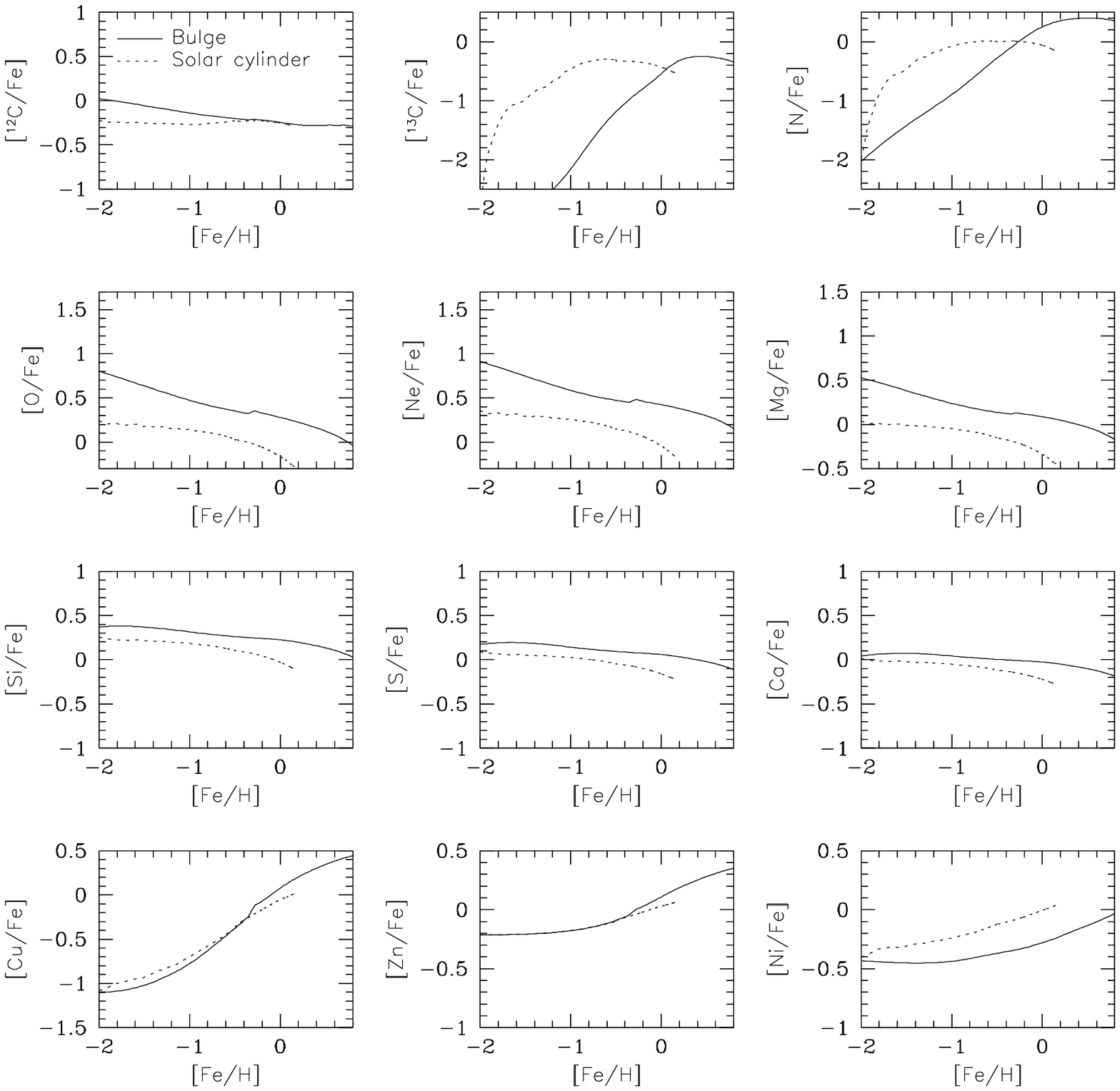}}
\hfill
\parbox[b]{\hsize}{
\label{fig6} 
\caption{Predicted [$\alpha$/Fe] vs. [Fe/H] relations for the bulge 
(continuous lines) and for the solar neighbourhood (dotted lines). The models
are from Matteucci et al. (1999).}}
\end{figure}

\section{Disks of Other Spirals}

{\it Abundance gradients}\par
Abundance gradients in dex/kpc
are known to exist also in the disk of other spirals
showing that they are
steeper in smaller disks, but this
correlation disappears if the gradients are expressed in units of 
dex/scalelength,
thus indicating the existence of a 
universal slope per unit scalelength (e.g. Garnett, 1998).
Another remarkable characteristic about abundance gradients in other 
spirals is that they appear to be
flatter in galaxies with central bars, suggesting that the dynamical 
effect of the bar can influence the evolution of the disk.

{\it The SFR}\par
The SFR is measured mainly from $H_{\alpha}$ emission (Kennicutt, 1998)
and implies a correlation with the total surface gas density (HI+$H_{2}$)
as discussed in section 2.1.

{\it Gas distributions}\par
Gas distributions, especially the HI distribution is known for 
a fair sample of spirals. There are indications of 
differences between field and cluster spirals (e.g. Skillman et al. 1996).

{\it Integrated colors}\par 
Studies of integrated colors of spiral disks reflect the observed abundance gradients and are well reproduced by an inside-out formation of disks similar to what assumed for the Milky Way
(Josey \&
Arimoto 1992; Jimenez et al. 1998; Prantzos \& Boissier 2000).

\section{Conclusions on the Milky Way and other spirals}
The comparison between observations and models for the Milky Way 
suggests that:

\begin{itemize}

\item The disk of the Galaxy formed mostly by infall of primordial or 
very metal poor gas accumulating
faster in the inner than in the outer regions (inside-out scenario,
i.e. $\tau_{D}(R)$ decreases with decreasing R).

\item In the framework of the inside-out scenario, the SFR should be 
a strongly varying function of the galactocentric distance.
In particular, it 
should either depend 
on the total surface mass density (feed-back mechanism)
or on the angular circular velocity of gas. Both formulations for the SFR 
are supported by observations. Under the assumption of a strongly 
varying SFR, the observed abundance gradients can be nicely reproduced.

\item Radial flows probably  are not the main cause of 
gradients but can help in reproducing the gas profile.

\item A constant  IMF (in time and space)
should be preferred, since variable IMFs have been tested and they 
cannot reproduce all of the observational constraints of the solar 
neighbourhood and the whole disk at the same time.

\item The properties of the disks of other spirals also 
indicate an inside-out disk formation.

\item
In the framework of semi-analytical models of galaxy formation 
(Mo, Mao \& White, 1998),
the evolution of galaxy disks can be described by means of 
scaling laws calibrated on the Galaxy with $V_c$ 
and $\lambda$ as parameters (Jimenez et al. 1998; 
Prantzos \& Boissier, 2000), where
$V_c$ is a measure of the mass of the dark halo
and $\lambda$ is a measure of the specific angular momentum 
of the halo.

\item Dynamical processes such as the formation of a central 
bar can influence
the
evolution of disks and deserve more attention in the future.

\item Abundance ratios (e.g. [$\alpha$/Fe])
in stars al large galactocentric distances can give us a clue to interpret
the formation of the disk and the halo (inside-out or outside-in). In fact, 
in a typical inside-out scenario we predict that 
the [$\alpha$/Fe] ratios would
decrease with galactocentric distance, due to the weaker and sometimes 
intermittent SF regime, whereas in an outside-in scenario we would expect 
the contrary due to the faster evolution of the more external region
which quickly consume the gas before Type Ia SNe have 
time to restore the bulk of Fe.

\item The predictions of chemical evolution models can be tested in a 
cosmological context to study the galaxy surface brightness and 
size evolution
as a function of redshift. Roche et al. (1998) have already 
done that and suggested that a size and luminosity evolution, 
as suggested by the inside-out scenario fits better the observations. 

\item Inside-out formation of the Galaxy is suggested also by the
fact that the globular clusters of the inner halo are coeval 
($\Delta t \sim 0.5$ Gyr, Rosenberg et al. 1999), 
whereas the age difference seems to increase in the outer halo. 

\item The metallicity distribution of stars in the Bulge as well
as the observed [$\alpha$/Fe] ratios
suggest a very fast formation of the Bulge ($\sim$ 0.5-0.8 Gyr),
due a the very fast SFR,
as predicted by succesfull chemical evolution models and 
by Elmegreen (1999).

\end{itemize}

\section{Elliptical Galaxies}
Elliptical galaxies are mainly found in galaxy clusters, they show a large range of luminosities and masses and are made by stars as old as those in globular clusters. No cold gas is observed in these galaxies
but hot X-ray gas halos are present.
Because of their large masses and metal content they are probably the most important factories of metals in the universe.

\subsection{Observational properties}
Here I recall the main observational features of elliptical galaxies:
\begin{itemize}

\item The existence of Color-Magnitude and 
color - velocity dispersion ($\sigma_o$) relations 
(colors become redder 
with increasing luminosity and mass; e.g. Bower et al. 1992)
is interpreted as a metallicity effect, namely as the fact that the 
metallicity decreases outwards similarly to what happens in spiral disks.

\item The existence of a $Mg_{2}$--$\sigma_o$ (where $Mg_{2}$
is a metallicity index) relation reinforces 
the previous point
(Bender et al. 1993; Bernardi et al. 1998; Colless et al. 1999; 
Kuntschner et al. 2001).

\item Inside ellipticals, 
$Mg_{2}$ correlates also with the escape velocity $v_{esc}$,
as first shown by  
Franx \& Illingworth (1990), indicating that
the magnesium index is larger where the escape velocity is larger.

\item $M/L_{B}$ increases by a factor of $\sim 3$ from faint 
to bright ellipticals implying a tilt of the fundamental plane
of ellipticals 
(Bender et al. 1992). The fundamental plane is the particular plane
occupied by these galaxies in the space defined by the stellar 
velocity dispersion,
the effective radius and the surface brigthness.

\item Abundance gradients inside ellipticals have been measured by means of
metallicity indices such as $Mg_{2}$ and $<Fe>$
(Carollo et al. 1993; Davies et al. 1993;
Kobayashi \& Arimoto 1999). These gradients correspond roughly 
to a gradient in [Fe/H] of the stellar component of
$\Delta [Fe/H] / \Delta log r 
\sim -0.3$.
\par
The average metallicity of the stellar component in ellipticals is
$<[Fe/H]>_{*} \sim -0.3$dex (from -0.8 to +0.3).

\item
By comparing synthetic metallicity indices with the observed ones
some authors 
(Worthey et al. 1992; Weiss et al. 1995; Kuntschner et al. 2001)
have suggested that the average stellar $<[Mg/Fe]>_{*}$ is larger than zero
(from 0.05 to + 0.3 dex) 
in nuclei of giant ellipticals. Moreover, there is indication that  
$<[Mg/Fe]>_{*}$ 
increases with increasing $\sigma_o$ ($M_{gal}$)
and luminosity (Worthey et al. 1992; Jorgensen 1999; 
Kuntschner  et al. 2001).\par
In particular the relation found by  Kuntschner et al. (2001) is:
[Mg/Fe]=$0.30(\pm 0.06) log \sigma_o - 0.52(\pm 0.15)$.

\end{itemize}

\subsection {Formation of Ellipticals}
Several mechanisms have been suggested for the formation and evolution of 
elliptical galaxies, in particular:
\begin{itemize}

\item Early monolithic collapse of a gas cloud or early merging
of lumps of gas where dissipation plays a fundamental role (Larson 1974;
Arimoto \& Yoshii 1987; Matteucci \& Tornamb\`e 1987 ). In this scenario
the star formation stops soon after a galactic wind 
develops and the galaxy evolves 
passively since then.

\item Bursts of star formation in merging subsystems made of gas
(Tinsley \& Larson 1979). 
In this picture star formation stops after the last burst and 
gas is lost via stripping or wind.

\item Early merging of lumps containing gas and stars in which 
some dissipation is present (Bender et al. 1993).

\item Merging of early formed stellar systems in a wide 
redshift range and preferentially at late epochs (Kauffmann et al. 1993).
A burst of star formation can occur during the major merging 
where $\sim 30\%$ of the stars can be formed (Kauffmann 1996).
\end{itemize}

The main difference between the monolithic collapse scenario and the 
hierarchical merging relies in the
time of galaxy formation, occurring quite early in the former scenario and
continuously in the latter scenario.
As we will see, there are arguments either in favour of the monolithic or
the hierarchical scenario. 

\subsection{Formation of Ellipticals at low z}

Here I recall some of the main arguments in favor of 
the formation of ellipticals 
at low redshifts:
\begin{itemize}

\item
Relative large values of the $H_{\beta}$ index measured in a sample of 
nearby ellipticals which could  indicate prolonged star formation 
activity up 
to 2 Gyr ago (Gonzalez 1993; Trager et al. 1998).

\item 
The tight relations in the fundamental plane are due to a conspiracy 
of age and metallicity in the sense that it should exist an 
age-metallicity anticorrelation 
implying that the more metal rich galaxies are also younger 
(Ferreras et al. 1999;
Trager et al. 2000).

\item The apparent paucity of high luminosity ellipticals 
at $z \sim 1$ compared to now claimed by a series of authors
(Kauffmann et al. 1993; 
Zepf, 1997;
Menanteau et al. 1999).
\end{itemize}

\subsection{Formation of Ellipticals at high z}
Here I recall the arguments in favor of a formation of ellipticals 
at high redshift:
\begin{itemize}

\item The tightness of the color-central velocity dispersion 
relation found for Virgo and Coma galaxies
(Bower et al. 1992). If the formation of ellipticals were a continuous 
process we should expect a much larger spread in the galaxy colors for a given
central velocity dispersion.
In particular, 
the argument goes like that:
from the observed color scatter one can derive $t_{H}-t_{F} \sim 2 $Gyr
(where $t_{H}$ is the Hubble time and $t_{F}$ the time of 
galaxy formation).
If $t_{H}=15$ Gyr then the youngest ellipticals  must have formed 
$\sim 13$ Gyr ago at $z \ge 2$ (Renzini, 1994).

\item The thinness of the fundamental plane for ellipticals in the
same two clusters, in particular the $M/L$ vs. $M$
relation  (Renzini  \& Ciotti 1993).

\item The tightness of the color-magnitude relation for ellipticals 
in clusters up to $z \sim 1$ (Kodama et al. 1998; Stanford et al. 1998)

\item The modest passive evolution measured for cluster 
ellipticals at intermediate redshift (van Dokkum \& Franx 1996; 
Bender et al. 1996).

\item Lyman-break galaxies at $z \ge 3$ 
where the SFR$\sim 50-100 M_{\odot}yr^{-1}$ could be the young ellipticals
(Steidel et al. 1996; 1998).

\item The strongly evolving population of Luminous Infrared Galaxies
suggesting that they are progenitors of massive spheroidals 
(Blain et al. 1999;
Elbaz et al. 1999).
\end{itemize}

\subsection{Models for ellipticals based on galactic winds}
Monolithic models assume that ellipticals suffer a 
strong star formation
and quickly produce galactic winds when the energy from SNe injected 
into the ISM equates the potential energy of the gas.
Then, star formation is assumed to halt after the development
of a galactic wind.
In this framework,
the evolution of ellipticals crucially depends on the time at which a 
galactic wind occurs, $t_{GW}$. For this reason, it is extremely important 
to understand the
SN feedback and star formation process.
The condition for the onset of a wind can be written as:
\begin{equation}
(E_{th})_{ISM} \ge E_{Bgas}
\end{equation}

The thermal energy of gas due to SNe and stellar wind heating is:
\begin{equation}
(E_{th})_{ISM}=E_{th_{SN}}+ E_{th_{w}}
\end{equation}

with 
\begin{equation}
E_{th_{SN}}= \int^{t}_{0}{\epsilon_{SN}R_{SN}(t^{`})dt^{`}}
\end{equation}

and 
\begin{equation}
E_{th_{w}}=\int^{t}_{0}\int^{100}_{12}{ \varphi(m) \psi(t^{`}) \epsilon_{w}dm
dt^{`}}
\end{equation}
for the contribution from SNe and stellar winds, respectively. 
The quantity $R_{SN}$ represents the SN rate (II and Ia).
The quantities
$\epsilon_{SN}= \eta_{SN}\epsilon_{o}$ with  $\epsilon_o=10^{51}$erg 
where $\epsilon_o$ is the typical SN blast wave energy , and
$\epsilon_{w}= \eta_{w}E_{w}$ with $E_{w}= 10^{49}$erg 
(typical energy injected by a $20M_{\odot}$ star taken as representative),
are the efficiencies for the energy transfer from SNe II and Ia into the ISM.
These efficiencies, $\eta_{w}$ and $\eta_{SN}$, can be assumed as 
free parameters or be calculated from the results of the evolution of 
a SN remnant in the ISM and the evolution of stellar winds. 
The SN feedback is, in fact, a crucial parameter.
The formulation of Cox (1972) for the efficiency of energy 
injection from SNe derives from following the evolution of the shock wave produced by the explosion into an ISM with constant density:

\begin{equation}
\epsilon _{SN}=0.72 \epsilon_o \,\, erg
\end{equation}
for $t_{SN} \le t_c=5.7 10^{4} \epsilon_o^{4/17} n_o^{-9/17}$ years,
where $t_c$ is the cooling time,
and
$t_{SN}$ is the time elapsed from the SN explosion, and:

\begin{equation}
\epsilon _{SN}= 2.2 \epsilon_o (t_{SN}/t_{c})^{-0.62}\,\, erg
\end{equation}
for $t_{SN} > t_{c}$.

With these prescriptions only few $\%$ of $\epsilon_o$
are deposited into the ISM (see Bradamante et al. 1998). 
However, multiple SN explosion should change 
the situation. Unfortunately, very few calculations 
of this type are available.
An important point to consider is also that
SNe Ia, which explode after Type II SNe, should provide more energy into 
the ISM than Type II SNe since they explode in an already formed cavity
(see Recchi, Matteucci \& D'Ercole, 2001).

The total mass of the galaxy is expressed as 
$M_{tot}(t)=M_{*}(t)+M_{gas}(t)+M_{dark}(t)$
with $M_L(t)=M_{*}(t)+M_{gas}(t)$
and the binding energy of gas is:

\begin{equation}
E_{Bgas}(t)=W_L(t)+W_{LD}(t)
\end{equation}
and:

\begin{equation}
W_L(t)=-0.5G{ M_{gas}(t) M_L(t) \over r_L}
\end{equation}
represents the potential well due to the luminous matter, whereas:
\begin{equation}
W_{LD}(t)= -Gw_{LD}{M_{gas}(t) M_{dark} \over r_L}
\end{equation}
is the potential well due to the interaction between dark and 
luminous matter,
where $w_{LD} \sim {1 \over 2\pi} S(1+1.37S)$
with $S= r_L/r_{D}$(Bertin et al. 1992).

The SFR is usually assumed to be:
\begin{equation}
SFR= \nu M_{gas}
\end{equation}
where $\nu \propto M_{L}^{-\gamma}$
($\gamma$ =-0.11, Arimoto  \& Yoshii 1987),
owing to the fact that the star formation efficiency is just 
the inverse of the timescale for star formation:
\begin{equation}
\nu= \tau_{SF}^{-1}
\end{equation}
and that:
\begin{equation}
\tau_{SF} \propto \tau_{coll} \propto \tau_{ff}
\end{equation}
with $\tau_{coll}$ and $\tau_{ff}$ being the collapse and the 
free-fall timescale, respectively.
Since dynamical timescales are longer for more massive galaxies, 
the efficiency of star formation should decrease with galactic mass.
The efficiency $\nu$ coupled with the increase of the potential well as
$M_{L}$ increases leads to the fact that more massive galaxies form 
stars for a 
longer period before suffering a galactic wind. This fact has been 
invoked for explaining the observed mass-metallicity relation (Larson, 1974).
However, this is  at variance with the observed
[Mg/Fe] vs. $\sigma_o$ relation,  since in this scenario the more massive 
ellipticals should show the lowest [Mg/Fe].

\subsection{Failure of Larson's Model}
There are different ways of obtaining that
$[Mg/Fe]$ in the stellar component increases when the galactic mass
$M_{L}$ increases.
These ways are:
\begin{itemize}
\item Different timescales for star formation (Worthey et al. 1992) 
in the sense that star formation should be more efficient
in more massive galaxies ($\nu \propto M_{L}^{\gamma}$).
In this case, the situation could be such that a
galactic wind occurs earlier in more massive systems, 
{\it the inverse wind scenario} (Matteucci 1994).

\item A  variable IMF from galaxy to galaxy favoring 
more massive stars (Mg producers)
in more massive galaxies (Worthey et al. 1992; Matteucci, 1994).

\item 
Different amounts and/or concentrations of dark matter as functions of 
$M_{L}$
In particular, less dark matter should be present
in the most massive systems (Matteucci, Ponzone \& Gibson, 1998). 
As a consequence 
of this, again
galactic winds occur earlier in more massive objects.

\item A selective loss of metals: more massive systems loose more Fe 
relative to $\alpha$-elements than less massive ones (Worthey et al. 1992).
\end{itemize}

As mentioned before, 
Matteucci (1994) proposed a model that she called 
{\it inverse wind model}, where the efficiency of star formation is an 
increasing function 
of galactic mass, thus implying a shorter period of star formation in 
massive ellipticals. In fact, the efficiency of star formation 
is chosen in such a way 
that in massive ellipticals the galactic wind occurs before than 
in less massive ones. This produces the increase of the [Mg/Fe] 
ratio as a function
of galactic mass.
This approach bears a resemblance with the merging scenario of 
Tinsley \& Larson
(1979) where the efficiency of star formation was assumed to increase 
with the total mass of the system. In the inverse wind model a very 
massive elliptical of $M_{L}=10^{12}M_{\odot}$ starts developing a wind 
before 
1 Gyr from the beginning of star formation, whereas a small ellipticals 
form stars for a longer period.
As a result, the average $<[Mg/Fe]> _{*}$
is larger in massive than in small ellipticals.
The same effect can be obtained by varying the IMF in such a way that more 
massive ellipticals tend to form more massive stars. It is worth noting
that this second hypothesis 
could also explain the tilt of the fundamental plane in $M/L_{B}$.
On the other hand, the hypothesis of the variable dark 
matter does not produce relevant effects, 
as shown by Matteucci et al. (1998).
The only assumption which has not been tested quantitatively is the 
selective loss of metals. The main problem with all of these alternatives is, 
in any case, the lack of a good physical justification depending on the 
poorly known
processes of star formation and SN feeback, and in the future some effort 
should be devoted to these fields.

It is worth noting that the hierarchical clustering scenario for galaxy 
formation cannot produce a solution to the observed [Mg/Fe] 
trend in ellipticals. 
In fact, it rather predicts the contrary, as shown by Thomas (1999) and 
Thomas et al. (2002). 
In fact, in the hierachical clustering the period of star formation in 
the most massive ellipticals is predicted to be the longest thus favoring low 
[Mg/Fe] ratios in massive objects, as shown in Figure 7.

\begin{figure}[t!]
\resizebox{\hsize}{!}{\includegraphics{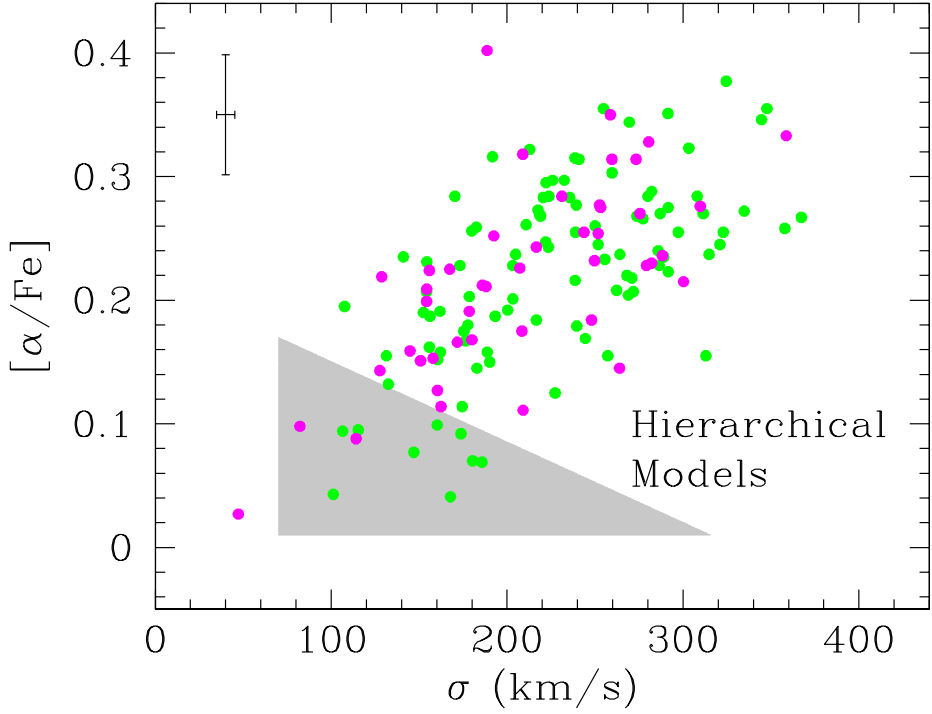}}
\hfill
\parbox[b]{\hsize}{
\label{fig7} 
\caption{Predicted [$\alpha$/Fe] vs. velocity dispersion 
for elliptical galaxies. Data are compared with predictions obtained 
by adopting
the star formation history assumed in hierarchical clustering models 
for galaxy formation.The models and the figure 
are from Thomas et al. (2002).}}

\end{figure}

\begin{figure}[t!]
\resizebox{\hsize}{!}{\includegraphics{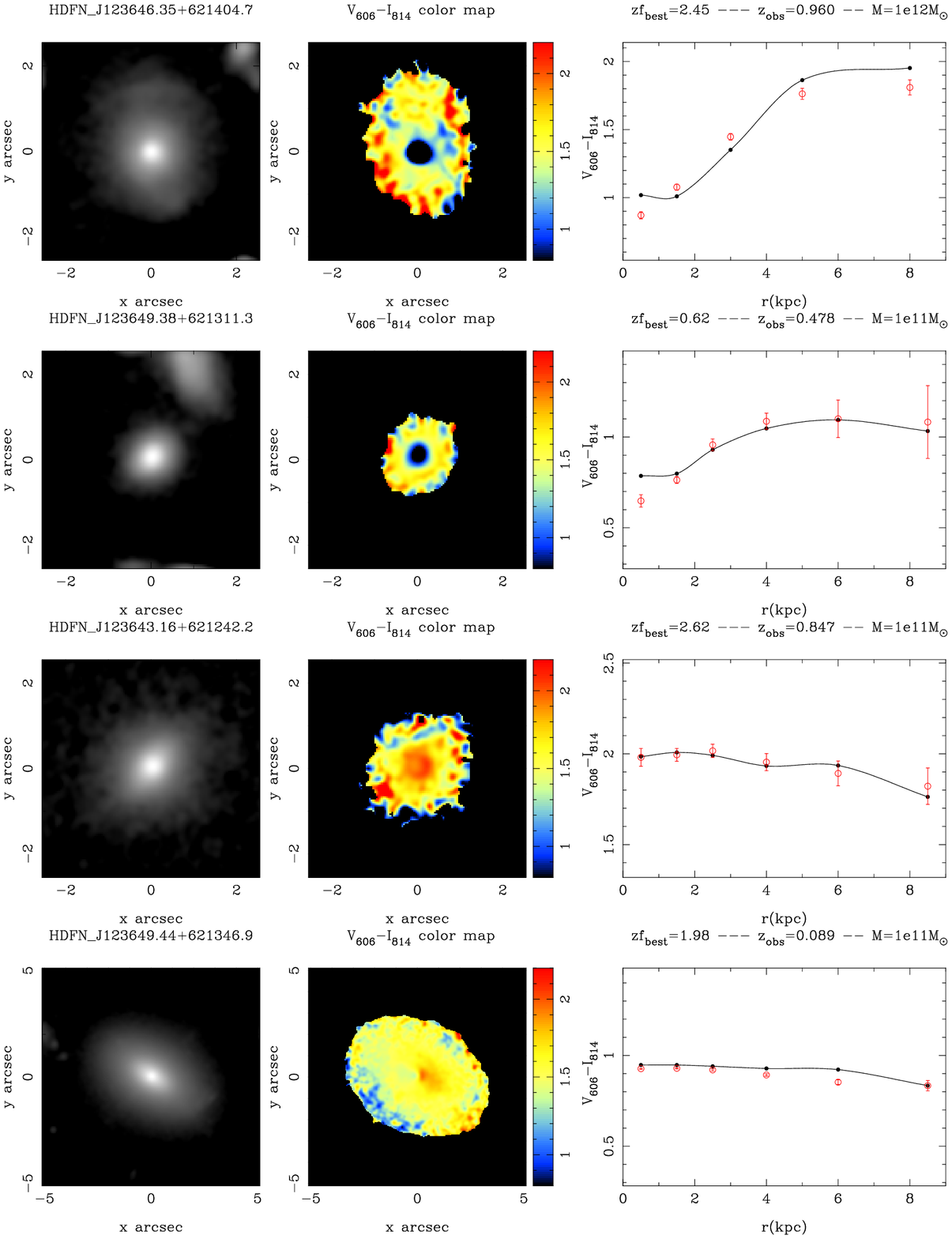}}
\hfill
\parbox[b]{\hsize}{
\label{fig8} 
\caption{Predicted and observed color gradients for some field
elliptical galaxies. From Menanteau et al. (2001).Figure shows 
from left to right: $I_{814}$-band surface brightness map, 
$V_{606}$-$I_{814}$ color pixel map and $V_{606}$-$I_{814}(r)$ 
color gradient. Open circles represent observed gradients while 
solid lines are the model predictions obtained by means of the 
Martinelli et al. (1998) model.
}}
\end{figure}

\subsection{Averaged Stellar Metallicities}
The metallicity in elliptical galaxies always refers to the metal 
content of the stars, in particular of the stellar population 
dominating in the visual light.
For this reason we should define the average stellar metallicity.
In particular, the average metallicity of a composite stellar population
averaged on the mass is:
\begin{equation}
<X_{i}>_{m}={1 \over S_1} \int^{S_1}_{0}{X_i(S) dS}
\end{equation}
where $1$ refers to the specific time $t_1$ and $S_1$ is the total mass of 
stars ever born.
This is the real average metallicity but in order to compare models and observations it is more appropriate to define the metallicity averaged on the visual
light.
In particular, the average metallicity of a composite stellar population 
averaged on the light is:
\begin{equation}
<X_{i}>_{L}={\sum_{ij}{n_{ij}X_iL_{Vj}} \over \sum_{ij}{n_{ij}L_{Vj}}}
\end{equation}
where $n_{ij}$ is the number of stars in the abundance interval $X_i$ and 
luminosity interval $L_{Vj}$.

It is worth noting that
$<X_{i}>_{m}$ is larger than $<X_{i}>_{L}$ for galaxies with 
$M_{L} < 10^{9}M_{\odot}$,
since metal poor giants dominate the visual light 
whereas they are similar for larger masses (Yoshii \& Arimoto 1987).

\subsection{Multi-Zone Models}
There are two multi-zone chemical evolution models for ellipticals  
available in the literature
(Martinelli et al. 1998;
Tantalo et al. 1998).
Martinelli et al. (1998) assumed that
the elliptical 
galaxy is divided in several concentric shells of thickness $\Delta R_i$. 
The binding energy of the gas in each shell is computed
after assuming a dark matter halo as described before. 
The model predicts that a galactic wind develops first in the 
external regions and then gradually in the more internal ones  
in agreement with the observed $Mg_{2}$ vs. $v_{esc}$ relation.
As a consequence of this, the 
star formation lasts for a shorter time in the external regions 
with the consequence
that an 
abundance gradient is created in agreement with observations. 
This model also predicts that we should observe higher
[Mg/Fe] ratios at larger galactocentric distances, the contrary of 
what should happens in the disks of spirals.
In this case, we can speak of {\it outside-in} formation. Unfortunately, 
the available data inside galaxies do not allow yet to observe such 
an effect. It is worth noting that this model can reproduce very well the 
observed color gradients in ellipticals, as shown in Figure 8.

\section{Conclusions on Ellipticals}
The main conclusions that we can draw from comparing theoretical models and 
observations for ellipticals can be summarized as follows:
\begin{itemize}

\item  In order to explain the observed $<[Mg/Fe]>_{*} > 0$
in giant ellipticals the dominant stellar population should have formed 
on a time scale no longer than 3-5 $\cdot 10^{8}$ yr, which corresponds
to the time at which the SNe Ia rate reaches a maximum in these systems 
with strong star formation.

\item  Uncertainties in the stellar yields of Fe and Mg from 
different authors can change the value of the [Mg/Fe] ratio but do not affect 
the conclusion above.

\item This observational finding 
argues against a hierarchical clustering formation scenario
and favors the fact that ellipticals, especially those in clusters, 
are mostly old systems (see also Menanteau et al. 2001).

\item   The increase of the [Mg/Fe] ratio with galactic mass
suggests either that more massive ellipticals are older 
systems or that the IMF
is not constant among ellipticals or both.

\item  Abundance gradients in ellipticals can be produced by 
biased winds and this would imply that [Mg/Fe] increases with increasing 
galactocentric distance.

\item  Better calibrations for metallicity indices are necessary, 
especially 
taking into account {non-solar ratios} in stellar tracks,
before drawing firm conclusions. 
\end{itemize}

\section{Evolution of Dwarf Galaxies}
Dwarf Irregular (DIG) and Blue Compact (BCG) galaxies
are very interesting objects for studying galaxy evolution since they 
are relatively unevolved objects (see Kunth \& \"Ostlin, 2000, for a 
recent exhaustive review).
In bottom-up cosmological scenarios they should be the
first self- gravitating systems to form, thus they
could also be important contributors to the population 
of systems giving rise
to QSO-absorption lines at high redshift (DLAs).
In general, they are rather simple objects with low metallicity
and large gas content, suggesting that they are either young or have
undergone discontinuous star formation activity (bursts).  
An important characteristic of these systems is
that they show a distinctive spread in their physical
properties, 
such as chemical abundances versus fraction of gas. 
Matteucci and Chiosi (1983) were among the first in studying the 
chemical evolution of dwarf galaxies. They adopted analytical models 
as those described in section 3 and showed that
closed-box models cannot account for the Z-log$\mu$ distribution
even if the number of bursts varies from galaxy to galaxy,
and suggested possible solutions to explain the observed spread:
\begin{itemize}
\item {a.} different IMFs

\item {b.} different amounts of galactic wind 

\item {c.} different amounts of infall

\end{itemize}
Later on, Matteucci and Tosi (1985) presented a numerical
model where galactic winds powered by SNe were taken into account.
They concluded that different wind rates from galaxy
to galaxy could explain the observed spread in O,N vs. log$\mu$ but not
the spread in the N/O vs. O/H diagram, suggesting that additional processes could have contributed to that.
For example, different amounts of primary N from galaxy to galaxy.
Kumai and Tosa (1992) suggested that different fractions
of dark matter in different objects could
explain the observed spread in the Z-log$\mu$ diagram.
Pilyugin (1993) forwarded the idea that the spread in 
the properties of these galaxies (i.e. He/H vs. O/H and N/O vs. O/H)
are due to self-pollution of the HII regions
coupled with ``enriched'' or ``differential'' galactic winds. Differential winds should carry out of the galaxy certain elements more than others, for example the metals ejected by SN II should be favored.
In more recent models (Marconi et al. 1994; 
Bradamante et al. 1998), the novelty was 
the contribution to the chemical enrichment of SNe of 
different type (II,Ia and Ib) together with differential winds.
In these papers the assumption was made that the products of SN II are lost 
more easily than the products of stars ending as WDs, 
such as C and N.
Larsen et al. (2001)  studied the chemical evolution of gas rich dwarf 
galaxies and concluded that  primary N production from massive stars is not 
necessary to reproduce the N/O vs O/H  and that the spread in this 
relation is 
likely to be due to the time-delay effect in the production of N relative 
to the production of O. 
In fact, in a starbursting regime when the starburst fades the elements 
produced by Type II SNe are no more produced whereas the elements produced on 
long timescales by single stars and Type Ia SNe are still produced. This 
causes the typical saw-tooth behaviour (see Figure 9). Therefore, the spread 
can be due to the fact that some objects are observed at different stages of 
the burst/interburts regime.
They also concluded that ordinary winds are better than enriched ones in 
reproducing the properties of these objects.
The existence of a luminosity-metallicity relation (although with spread)
can be an indication for galactic winds acting more efficiently in low 
mass than in high mass high potential well objects.

\subsection{Evidences for Galactic Winds}
Meurer et al (1992), Papaderos et al. (1994), Lequeux et al. (1995) and
Marlowe et al. (1995) all suggested the existence of galactic winds 
in dwarf starbursting galaxies.
The evidence is gathered by the indication of outflowing material 
travelling at a speed larger than the assumed mass of the objects.
Papaderos et al. estimated a galactic wind flowing at a 
velocity of 1320 Km/sec
for VIIZw403. The escape velocity estimated for
this galaxy being 50 Km/sec.
Lequeux et al.(1995) suggested a galactic wind in Haro2=MKn33 flowing at 
a velocity of $\simeq 200 Km/sec$.
Martin (1996;1998) found supershells in 12 dwarfs including
IZw18 and concluded that they imply an outflow 
which in some cases can become a wind 
(namely the material is lost from the galaxy).

Recent chemo-dynamical simulations for one instantaneous starburst 
also suggest the possibiliy of galactic winds and that these winds are
metal enriched (MacLow \& Ferrara 1999; Recchi et al. 2001).

\subsection {Results for BCG from chemical models}
Purely chemical models (no dynamics) have been computed by several authors
by varying the number of bursts, the time of occurrence of bursts, 
$t_{burst}$, the
star formation efficiency, the type of galactic wind, the IMF and the 
nucleosynthesis prescriptions 
(Marconi et al. 1994; 
Kunth et al.1995; Bradamante et al.1998).
The main conclusions of these papers can be summarized as follows:

\begin{itemize}
\item 
The number of bursts should be 
$N_{bursts} \le 10$, the star formation efficiency should vary from 0.1 to 
0.7 $Gyr^{-1}$ for either Salpeter or Scalo (1986) IMF but Salpeter IMF 
is favored.

\item Enriched winds, carrying out material at a rate proportional 
to the star formation rate, seem to be preferred (but see Larsen et al. 2001).
The observed scatter in the observational properties can be due either to 
the winds or to the delay in the production of different elements.

\item If the burst duration is relatively short (no more than 100 Myr),
SNe II dominate the chemical evolution and energetics of 
starburst galaxies, while stellar winds seem to be  negligible 
after the onset of SNe II.

\item The [O/Fe] ratios tend to
be overabundant  
due to the 
predominance of Type II SNe during the bursts. Models with a large 
number of bursts $N_{burst}$=10 - 15 can give 
negative [O/Fe]. 
\end{itemize}

\begin{figure}[t!]
\resizebox{\hsize}{!}{\includegraphics{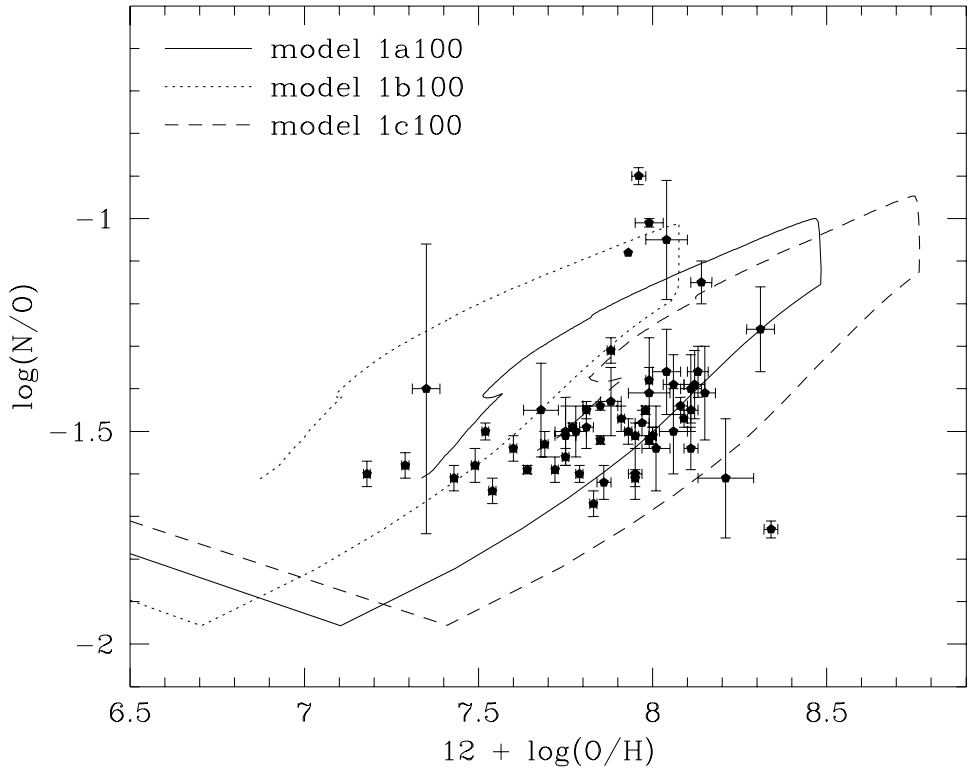}}
\hfill
\parbox[b]{\hsize}{
\label{fig9} 
\caption{The log(N/O) vs. 12 +log(O/H) for a sample of BCG. The data are 
from Recchi (2002). Overimposed are three models with a single 
burst of star formation and different star formation efficiency. 
In particular,
the dotted line corresponds to an efficiency $\nu=1 Gyr^{-1}$, 
the continuous line to
$\nu=2.5 Gyr^{-1}$ and the dashed line to $\nu=5Gyr^{-1}$.
The burst duration is 100 Myr.
As one can see, the saw-tooth behaviour typical of a bursting 
mode of star fotmation is evident. 
}}
\end{figure}

\begin{figure}[t!]
\resizebox{\hsize}{!}{\includegraphics{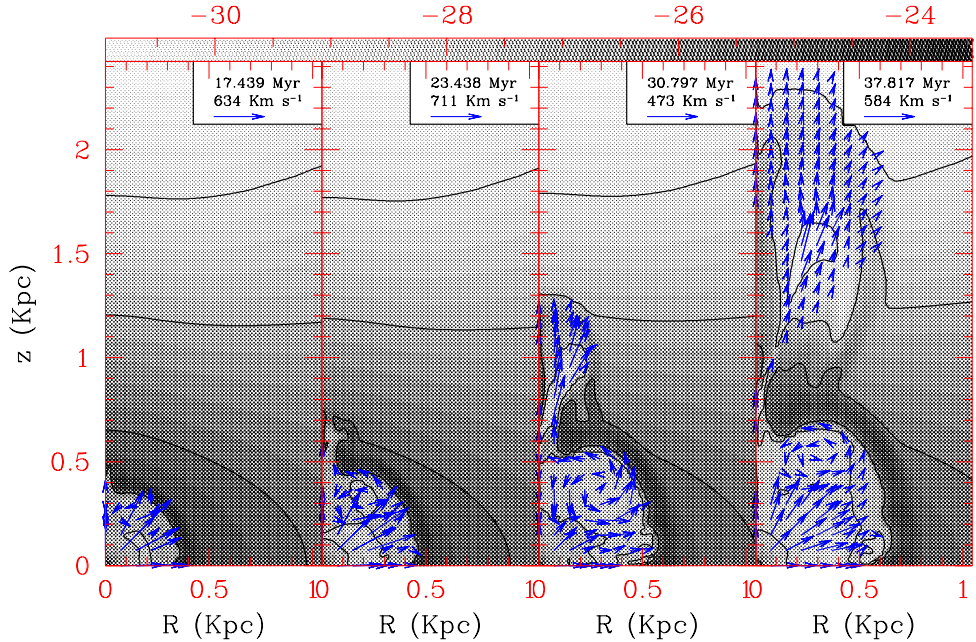}}
\hfill
\parbox[b]{\hsize}{
\label{fig10} 
\caption{Chemo-dynamical simulation of an instantaneous starburst
from Recchi et al. (2001). The development of a galactic wind is evident 
along the z-axis, since the assumed configuration is flattened.The age of 
the burst is indicated in Myr.}}
\end{figure}

\subsection{Results from chemo-dynamical models}

Recent chemo-dynamical models (Recchi et al. 2001;2002)
assuming an instantaneous starburst but following in great detail 
the evolution of several chemical elements (H, He, C, N, O, Mg, Si and Fe)
and adopting an efficiency for SN II energy transfer $\eta_{SNII}$=0.003
and for SN Ia $\eta_{SNIa}$=1.0,
suggest that
the starburst triggers indeed a galactic wind (see Figure 10). 
In particular, 
the metals leave the galaxy more 
easily than the unprocessed gas and among the metals the
SN Ia ejecta leave the galaxy more easily than the SN II ejecta. This is 
due to the assumed efficiencies for energy transfer for the two types of SNe. 
This assumption is in turn based on the fact that Type Ia SNe explode in an 
already heated and rarified medium (thanks to the Type II SNe which 
explode first) 
and therefore can transfer all of their energy into the ISM.
As a consequence of this type of evolution the following conclusions 
can be drawn:
\begin{itemize}

\item A selective loss of metals seems to occur in dwarf gas-rich galaxies.

\item As a consequence of the selective winds, the [$\alpha$/Fe] ratios
inside the galaxy are predicted to be larger  than the [$\alpha$/Fe] 
ratios outside the galaxy. In fact, the products of SNe Ia are lost 
more efficiently than those of SN II.

\item At variance with previous studies, most of the metals 
are already in the cold gas phase after 8-10 Myr owing to the fact that the 
superbubble
does 
not break immediately (the SNe II inject only a fraction of their initial 
blast wave energy into the ISM) and thermal conduction can act efficiently.

\item  The model well reproduces the properties of IZw18 (the most metal  
poor galaxy known locally) if two
bursts are assumed and are separated by 300 Myr interval.
\end{itemize}

\subsection{Dwarf galaxies and DLA Systems}       
Finally, before concluding this section, we would like to draw the attention 
upon the fact that there are similarities between BCG, DIG and DLAs or more 
in general between DLAs and systems with a low level of star formation. 

The nature of DLA systems is under debate and the
abundance ratios measured there 
can be used as  a diagnostic to infer their  nature and age.
Matteucci et al. (1997) suggested that some DLA showing high N/O
ratio could be BCG suffering selective winds. 
In this respect, the similarity between DLAs
and IZw18 may suggest that IZw18 is a survivor proto-galaxy which has 
just started forming stars.
If this is true, dwarf irregular galaxies should be born at any time during 
the age of the universe, unlike elliptical galaxies which appear 
to have formed 
a long time ago and in a short time interval (see section 9).

\begin{figure}[t!]
\resizebox{\hsize}{!}{\includegraphics{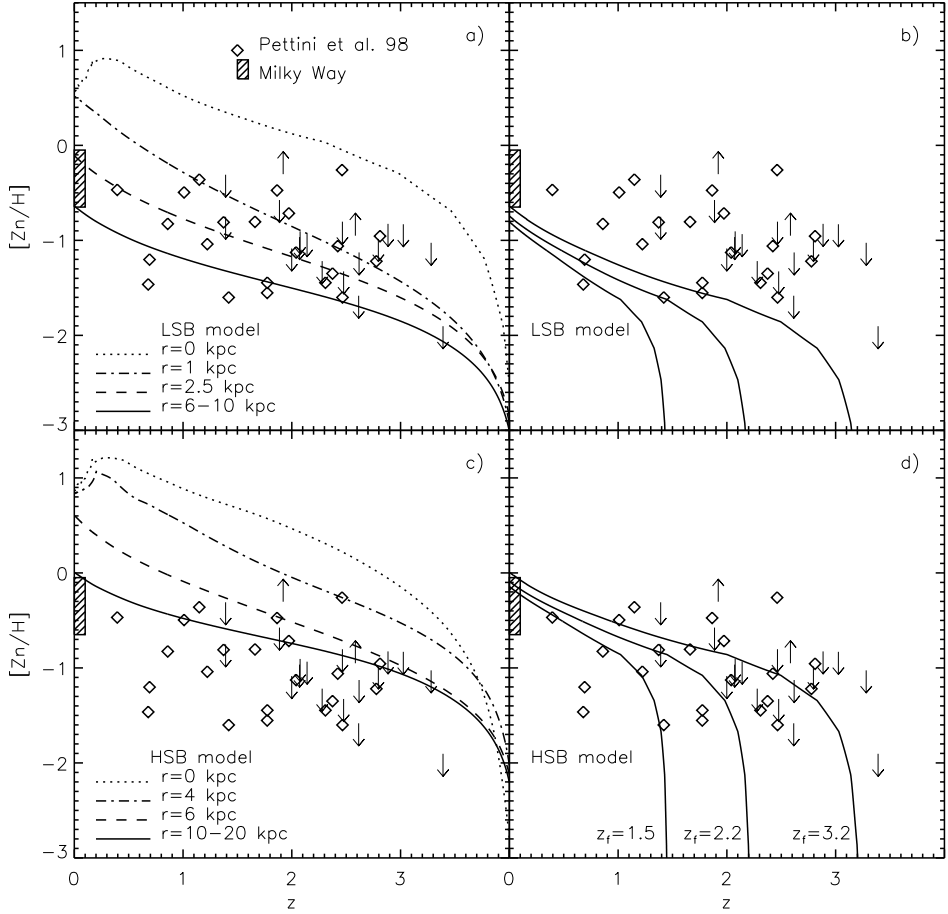}}
\hfill
\parbox[b]{\hsize}{
\label{fig11} 
\caption{Redshift evolution of [Zn/H] for high surface brightness spirals 
(HSB) (bottom panels) and low surface brightness  spirals (LSB) (upper panels)
for different galactocentric distances in the disk. a) The evolution 
in metallicity of LSB formed at z=4 fits well the DLA values measured 
by Pettini et al. (1999).b) If LSB disks formed late, however, then they 
do not fit the data.
c) Conversely, HSB disks that formed at z=4 become metal rich too quickly 
to explain the observations; d) only if there is a continuously forming 
population
of HSB disks between z=4 and z=1 then they can account for the metallicities 
of the DLAs.}}
\end{figure}

Plots of [$\alpha$/Fe] vs. [Fe/H] 
and plots of [$\alpha$/Fe] vs. redshift 
should be used to infer the nature  and the age
of these objects, when compared with chemical evolution predictions.

The observed abundance ratios 
seem to indicate that DLAs show almost solar [$\alpha$/Fe] ratios at low [Fe/H] (Pettini et al. 1999; Centurion et al. 2000). This is an indication that they are objects where the star formation proceeded slowly and that they have probably started to form stars long before the redshift at which 
we observe them. 
This could indicate that we are
looking at 
the external regions of disks or at dwarf starbursting 
systems in the interburst 
phases (see Figure 11).
However, more data are necessary to assess this point.
One common problem related to the measurement of the abundances in 
DLAs is that some elements are dust depleted such as Si and Fe and therefore 
one should try to observe non-refractory elements such as N, O, S and Zn.
One should also remember that Si and Ca do not strongly behave 
as $\alpha$-elements since they are produced 
in a non-negligible way in SNe Ia.

\section{Chemical Enrichment of the ICM}       
After having discussed the chemical evolution of galaxies it is important 
to conclude by studying how the evolution of galaxies can affect the 
chemical enrichment of the intracluster (ICM)  and intergalactic medium.
In the past years a great deal of work has been presented on the subject.
The first work on chemical enrichment of the ICM was
by Gunn \& Gott (1972), Larson \&
Dinerstein (1975), Vigroux (1977), Himmes \& Biermann (1988).
In the following years,
Matteucci \& Vettolani (1988) started a more detailed 
approach to the problem followed by David et al. (1991), 
Arnaud et al. (1992), 
Renzini et al. (1993), 
and many others.
The majority of these papers assumed that galactic winds  
(mainly from ellipticals) are responsible for the ICM chemical enrichment.
Alternatively, the abundances in the ICM could be due to ram 
pressure stripping (Himmes \& Biermann 1988) or to pre-galactic Population III
stars (White \& Rees 1978).

\subsection{Models for the ICM}
Here we will describe briefly the metodology developed by 
Matteucci \& Vettolani (1988) (hereafter MV88).
Starting from SN driven galactic wind models for ellipticals (see section 9)
they computed the ejected masses vs. final baryonic total galactic masses for
galaxies of different initial luminous masses (from $10^{9}$ to 
$10^{12}M_{\odot}$):

\begin{equation}
M_{i}^{ej}=E_iM_{f}^{\beta_i},
\end{equation}
where $i$ refers to either a single chemical element or the total gas.
Then, they integrated $M_{i}^{ej}$ over the cluster mass 
function obtained from the
Schechter (1986) luminosity function (LF) by assuming that only E and S0
galaxies contribute to the 
chemical enrichment. This assumption 
was later confirmed by data of Arnaud et al.  (1992).
The total masses ejected into the ICM in the form of 
single species $i$ and total gas are:

\begin{equation}
M_{i,clus}^{ej}(>M_{f})=E_i f n^{*}(h^{2}k)^{\beta_i} 
10^{-0.4 \beta_i(M^{*}-5.48)}\cdot
\Gamma[(\alpha +1 + \beta_i), 
(M_f^{*} h^{2}/k) 10^{-0.4(M_B^{*}-5.48)}] 
\end{equation}
where $\alpha$ is the slope of the LF,
$f$ is fraction of ellipticals plus S0, 
$M^{*}$ is the mass at the ``break'' of the LF and
$M_B^{*}$ is the magnitude at the break.

\subsection{MV88 Results}
MV88 considered 4 galaxy clusters: Perseus, A2199, Coma and Virgo 
for which $f$, $n^{*}$,
$M^{*}$, $M_{Fe}$ and $M_{gas}$ were known.  In Table 1 we show the 
results they obtained for the 4 clusters concerning the 
total masses ejected by all galaxies in the clusters in the form of total 
gas and Fe, compared with the observed ones. It is immediate to see 
from Table 1 
that their model could reproduce well the total amount of Fe but they 
failed in reproducing the total gas mass. However, this is not a failure 
of the model but the indication that most of the gas in clusters has a 
primordial origin, namely has never been processed inside stars. The 
same conclusion was reached later by David et al. (1991) and 
Renzini et al. (1993) among others, but see Chiosi (2000).

\begin{table}
\centering
\paragraph{ Table 1. Predicted and observed quantities}
\begin{tabular}{c c c c c} 
\multicolumn{5}{c}{}\\
Cluster & $(M_{Fe})_{pred}$ & $(M_{Fe})_{obs}$ &  $(M_{gas})_{pred}$ & $(M_{gas})_{obs}$\\
\hline\noalign{\smallskip}
Perseus  & $2.6 10^{11}$ & $1.9 10^{11}$ & $2.0 10^{13}$ & $3.0 10^{14}$\\
A2199    & $1.3 10^{11}$ & $1.2 10^{11}$ & $10^{13}$ & $1.5 10^{14}$\\
Coma     & $2.3 10^{11}$ & $3.1 10^{11}$ & $1.8 10^{13}$ & $1.5 10^{14}$\\
Virgo    & $1.4 10^{10}$ & $1.6 10^{10}$ & $10^{12}$ & $2.0 10^{13}$\\
\end{tabular}
\end{table}

\begin{table}
\centering
\paragraph{ Table 2. Predicted Fe abundances and abundance ratios}
\begin{tabular}{c c c c } 
\multicolumn{4}{c}{}\\
\hline
\noalign{\smallskip}
Cluster & $X_{Fe}/X_{Fe_{\odot}}$ & [Mg/Fe] &  [Si/Fe]\\
\hline\noalign{\smallskip}
Perseus & 0.65 & -0.65 & -0.080\\
A2199   & 0.65 & -0.60 & -0.096\\
Coma    & 0.43 & -0.63 & -0.110\\
Virgo   & 0.53 & -0.64 & -0.090\\
\end{tabular}
\end{table}

Therefore, the
$X_{Fe}/X_{Fe_{\odot}}$ shown in Table 2 was calculated as 
$(M_{Fe})_{pred}/(M_{gas})_{obs}$. The predicted
Fe abundance in the ICM relative to the Sun is
in agreement with the observations 
then and now $(X_{Fe}/(X_{Fe_{\odot}})_{obs}=0.3-0.5$
(Rothenflug \& Arnaud 1985; White 2000).
Low values for [Mg/Fe] and [Si/Fe] were predicted due to the 
assumption that all the Fe,
produced 
by Type Ia SNe, was soon or later ejected into the ICM,
as shown in Table 2. With Salpeter IMF, Type Ia SNe 
contribute $ \ge 50\%$
of the total Fe.

\subsection{[$\alpha$/Fe] Ratios in the ICM}
Models including Type Ia and II SNe predict  an asymmetry in the 
[$\alpha$/Fe] ratios
($> 0$ inside the ellipticals and $< 0$ in the ICM) due to the 
different roles of SNe II and Ia in Fe production (Renzini et al. 1993.)
ASCA results (Mutshotzky et al. 1996) originally
suggested [$\alpha$/Fe]$_{ICM} >$ 0
(+0.2 dex). However, Ishimaru \& Arimoto (1999)  
pointed out that [$\alpha$ /Fe] $\sim$ 0
in the ICM if the meteoritic Fe abundance 
is adopted instead of the photospheric value adopted in the other paper.
On the basis of the ASCA results on the overabundance of the 
$\alpha$-elements,  Matteucci \& Gibson (1995) discussed how to reproduce 
the [$\alpha$/Fe]$>$0 ratios both in stars and ICM.  
They adopted the same  SN feedback as MV88 and dark matter halos 
were also included in the 
galaxy models. They concluded that 
it is possible to obtain overabundances of the $\alpha$-elements in the ICM
only if 
not all of the produced Fe is lost from the galaxies
(i.e. only early winds). 
On the other hand, MV88 had assumed that all the Fe is soon or later 
ejected into the ICM.
They concluded that a flat IMF (x=0.95) 
plus only early winds can reproduce
[$\alpha$/Fe]$_{ICM} >0$.
A similar conclusion was reached by Loewenstein \& Mutshotzky (1996).
However, the situation is not very realistic since the gas in cluster 
galaxies is likely 
to be stripped soon or later because of environmental effects,
and the [$\alpha$/Fe] ratios in the ICM are likely to be solar or undersolar.
More recently,
Martinelli et al. (2000) recomputed the ICM enrichment
by adopting a more realistic model for the evolution of ellipticals 
(multi-zone).
The model (already described in section 9) predicts a more extended 
period of galactic wind and
more metals and gas in the ICM than the one-zone model 
with only early winds.
They computed the evolution of abundances vs. redshift 
for a constant LF and concluded
that there is no evolution between z=1 and z=0 and [$\alpha$/Fe]$_{ICM}
\le$0 at the present time.
Very recently Pipino et al. (2002) computed the chemical 
enrichment of the ICM as a function of redshift by considering the evolution 
of the cluster luminosity function and confirmed the conclusions of 
Martinelli et al. (2000). Very recent data from XMM-Newton 
(Gastaldello \& Molendi, 2002)
seem to indicate a low [O/Fe]$<0$ ratio for the ICM, thus supporting 
models where Type Ia SNe are efficient and all the produced 
Fe is ejected into the ICM.
In Figure 12 we show the recent results of Pipino et al. (2002). 
In the upper panel of Fig. 12 is indicated the predicted evolution as a 
function of redshift of 
the thermal energy per particle in the ICM
due to galactic winds from E and S0 galaxies.In the lower panel is shown the 
evolution of the total mass of Fe ejected by the cluster galaxies into the ICM
as a function of redshift. It is worth noting that 1 keV per particle 
is the energy required to explain the observed $L_{X}-T$ relation in 
clusters (e.g. Borgani et al. 2001). The model in Figure 12 shows that the 
energy
injected by SNe is not enough (at maximum 0.4 keV) 
and that other sources of energy are required, 
such as active galactic nuclei and QSO. Figure 12 also shows how important 
is the contribution from Type Ia SNe to the chemical enrichment and  
energetic content of the ICM. In fact, SNe II are important in triggering 
the galactic wind but after star formation stops the wind is sustained 
only by Type Ia SNe.

\subsection{[$\alpha$/Fe] ratios and IMLR}

Abundance ratios and the Iron Mass to Light Ratio (IMLR) are good tests 
for the evolution of galaxies in clusters since they 
do not depend on the total 
cluster gas mass.
The IMLR is defined as (Renzini et al. 1993):
\begin{equation}
IMLR =M_{Fe}^{ICM}/L_B 
\end{equation}
and the IMLR  $\propto h^{-0.5}$.
For $ H_{0}=50$ the IMLR=0.02 ($M_{\odot}/L_{\odot}$) 
(Arnaud et al. 1992) practically constant among rich 
clusters but it drops for poor clusters and groups.
On the one hand, the IMLR can impose constraints on the 
IMF in cluster galaxies, 
on the baryonic history of the clusters/groups and on the number of 
SNe exploded in galaxies. In fact, the most straightforward interpretation 
of the constancy of the IMLR among rich clusters is that they did not loose Fe
at variance with the groups or small clusters having a lower IMLR indicating 
a loss of baryons (Renzini, 1997).
On the other hand, the
[$\alpha$/Fe] ratios can impose constraints on stellar nucleosynthesis, 
IMF, different roles of SNe and SN feedback.

\begin{figure}[t!]
\resizebox{\hsize}{!}{\includegraphics{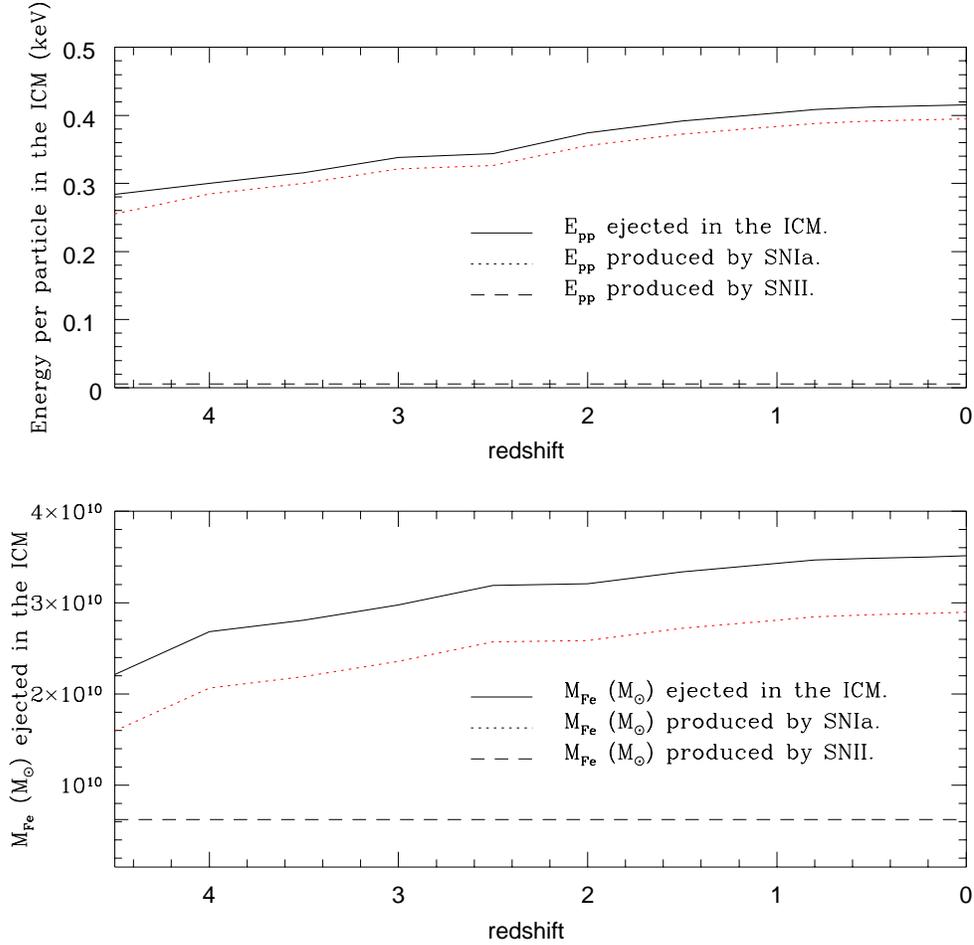}}
\hfill
\parbox[b]{\hsize}{
\label{fig12} 
\caption{The predicted evolution of the thermal content of the ICM 
(upper panel) as a function of redshift. The different contributions from SNe 
of different type
is indicated. In the lower panel is shown the evolution of the total Fe 
mass as a function of redshift. The assumed cosmological model is 
$\Omega_{m}=0.3$, $\Omega_{\lambda}=0.7$ and h=0.70. 
}}
\end{figure}

\section{Conclusions on the ICM}
The main conclusions on the chemical enrichment of the ICM can be 
summarized as follows:
\begin{itemize}

\item 
Good models for the chemical enrichment of the ICM should 
reproduce at the same time the 
[$\alpha$/Fe] ratios inside galaxies and in the ICM.
They should also reproduce the IMLR.

\item
Crucial parameters are:
SN feedback,
SN nucleosynthesis, 
IMF and
whether all the stellar ejecta (produced over a Hubble time) can reach the ICM
or remain bound to the parent galaxy.

\item  Type Ia SNe play a fundamental role in the chemical and 
energy content evolution of the ICM. 

\item
Good models for the ICM enrichment, assuming that the energy transfer from 
SN II is $\sim 3\%$ and from SNIa is 100$\%$, predict an energy per particle 
in the ICM of $E_{p} \sim 0.4$ keV, not enough
to break the self- similarity in clusters. 
\end{itemize}

\end{document}